\documentclass[12pt]{article}
\pdfoutput=1
\usepackage{color,graphicx,subfigure,amsmath,epsfig,amsfonts,amsthm}
\usepackage[margin=1in]{geometry}
\usepackage[super,sort&compress]{natbib}
\DeclareGraphicsExtensions{.png,.jpg,.pdf}

\newcommand{\bv}{{\bf b}}
\newcommand{\xv}{{\bf x}}
\newcommand{\hv}{{\bf h}}
\newcommand{\vect}[1]{\boldsymbol{#1}}
\newcommand{\mean}[1]{\langle #1 \rangle}

\usepackage{authblk}

\title{Associative content-addressable networks with exponentially many robust stable states}
\author[1]{Rishidev Chaudhuri}
\author[1]{Ila Fiete}
\affil[1]{Center for Learning and Memory and Department of Neuroscience, 
\protect\\ 
The University
of Texas at Austin,
Austin, Texas, USA.}
\date{}

\begin{document} 
\maketitle 

\begin{abstract}
The brain must robustly store a large number of memories, corresponding to the many events encountered over a lifetime. However, the number of memory states in existing neural network models either grows weakly with network size or recall fails catastrophically with vanishingly little noise. We construct an associative content-addressable memory with exponentially many stable states and robust error-correction. The network possesses expander graph connectivity on a restricted Boltzmann machine architecture. The expansion property allows simple neural network dynamics to perform at par with modern error-correcting codes. Appropriate networks can be constructed with sparse random connections, glomerular nodes, and associative learning using low dynamic-range weights. Thus, sparse quasi-random structures---characteristic of important error-correcting codes---may provide for high-performance computation in artificial neural networks and the brain.
\end{abstract}

\section*{Introduction}
Neural long-term memory systems have high capacity, by which we mean that the number of memory states is large. Such systems are also able to recover the correct memory state from partial or noisy cues, the definition of an associative memory. If the memory state can be addressed by its content, it is furthermore called content-addressable. 

Classic studies of the dynamics and capacity of associative content-addressable neural memory (ACAM) have focused on connectionist neural network models commonly called Hopfield networks \cite{Little74, Hopfield82, Grossberg88}, which provide a powerful conceptual framework for thinking about pattern completion and associative memory in the brain. Here we continue in this tradition and examine constructions of ACAM networks in the form of Hopfield networks and their stochastic equivalents, Boltzmann machines. We show that it is possible to construct associative content-addressable memory networks with an unprecedented combination of robustness and capacity.  

Hopfield networks consist of binary nodes (or ``neurons'') connected by symmetric (undirected) weights (Figure 1a). At each time-step, one neuron updates its state by summing its inputs; it turns {\em on} if the sum exceeds a threshold. All neurons are updated in this way, in random sequence. The network dynamics lead it to a stable fixed point, determined by the connection strengths and initial state (Figure 1b). Equivalently, stable fixed points may be viewed as minima of a generalized energy function. Any state in the basin of attraction of an energy minimum flows to the minimum as the network dynamics push the system downhill in energy (Figure 1c). The set of stable fixed points, fully determined by the network weights, are the memory states. They are robust if their basins are sufficiently large. The Boltzmann machine \cite{Hinton83} is a stochastic version of the Hopfield network: Each neuron becomes active with a probability proportional to how much its summed input exceeds threshold, and the network dynamics approach and remain in the vicinity of stable fixed points of the corresponding deterministic dynamics. Hopfield networks, Boltzmann machines, and related constructions such as autoencoders have proved to be versatile statistical models for natural stimuli and other complex inputs \cite{Olshausen96,Tishby99,Hinton06,Hillar14images, LeCun15}.

Very generally, in any representational system there is a tradeoff between capacity and noise-robustness \cite{McEliece77, Mackay04}: a robust system must have redundancy to recover from noise, but the redundancy comes at the price of fewer representational states, Figure 1d. Memory systems exhibit the same tradeoff.

A network consisting of $N$ binary nodes has at most $2^N$ states; memory states are the subset of stable states. We define a {\em high-capacity} memory network as one with exponentially many stable states as a function of network size (i.e., $C \sim e^{\alpha N}$, for some constant $0<\alpha\leq1$). A high-capacity network retains a non-vanishing {\em information rate} $\rho$ ($\rho \equiv {\log(C)}/N = \alpha$) even as the network grows in size. 
 
A {\em robust} system is one that can tolerate a small but constant error rate ($p$) in each node, and thus a linearly growing number of total errors ($pN$) with network size. Tolerance to or robustness against such errors means that the network dynamics can still recover the original state. For robustness to growing numbers of errors with network size, each memory state must therefore be surrounded by an attracting basin that grows with network size. 

In sum, high-capacity memory networks must have the same order of memory states as total possible states, both growing exponentially with network size; at the same time, the basins around each memory state must grow with network size. Can these competing requirements be simultaneously satisfied in any network? 

Hopfield networks with $N$ nodes and $\sim N^2$ pairwise analog (infinite-precision) weights can be trained with simple learning rules to learn and exactly correct up to $N/(2 \log(N))$ random binary inputs of length $N$ each \cite{Mceliece87} or imperfectly recall $0.14N$ states (with residual errors in a small fraction of nodes) \cite{Amit85}. With sparse inputs or better learning rules, it is possible to store and robustly correct $\sim N$ states \cite{Kanter87, Tsodyks88,Hillar12,Alemi15}. Independent of learning rule, the capacity of the Hopfield networks is theoretically bounded at $\sim N$ arbitrary states \cite{Abu85, Gardner88, Treves91}.

If {\em higher-order} connections are permitted (for instance, a third order weight connects neurons $i, j,$ and $k$ with each other), Hopfield networks can robustly store $C \sim N^{K-1}$ memories \cite{Baldi87, Burshtein98}, where $K$ is the order of the weights (thus there are up to $N^K$ weights). This capacity is still polynomial in network size.

\begin{figure}
    \begin{center}
        \includegraphics[width=\textwidth]{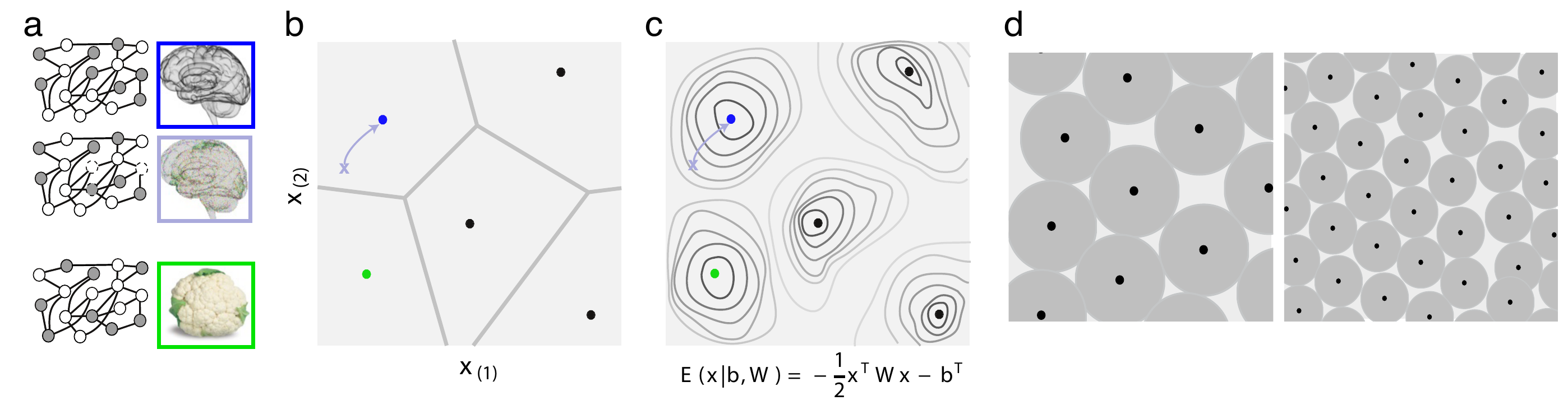}
    \end{center}
    \caption{{\bf Hopfield networks and the robustness-capacity tradeoff.} (a) Left: Schematic of different activity states in a Hopfield network (black lines: symmetric connections; dark nodes: inactive; white: active).  Right: the different activity states may be viewed as corresponding to  different patterns. Top and bottom depict states that are stable states of the network dynamics. Center: an intermediate state that is not a stable state. (b) Schematic state-space view of the Hopfield network. Any point in state-space corresponds to some state of the network. The dots represent stable states; the blue and green points represent the stable patterns from (a), while the lavender cross represents the intermediate state. Different stable states are separated from each other and points in the neighborhood of a stable state flow to it under the network's dynamics (e.g., the lavender cross to the blue stable state). (c) An energy landscape view: The stable states have low energy (dots) and correspond to minima in a rugged energy landscape. Different minima are separated by energetic ridges. The network dynamics corresponds to a descent on the energy landscape to a local minimum or stable state.  (d) Fundamental tradeoff between robustness and capacity: Each stable state is surrounded by a ``basin of attraction'' consisting of nearby states that flow to it (left). If these are viewed as noisy or corrupted copies of the stable state, the network is a decoder that performs denoising of perturbed versions of its stable (memory) states. Packing in more stable states (right) reduces the robustness of the network to noise.}
\label{fig:eccs_and_hopf_nets}
\end{figure}

Spin glasses (random-weight fully recurrent symmetric Hopfield networks) possess exponentially many local minima or (quasi)stable fixed points \cite{Tanaka80, McEliece85, Baldi87, Sourlas89}, Hopfield networks designed to solve constraint satisfaction problems exhibit $\sim 2^{\sqrt{N}}$ stable states \cite{Hopfield85, Hopfield86, Tank86}, and a recent construction with hidden nodes shows capacity that is exponential in the ratio of the number of hidden to input neurons\cite{Alemi17}. However, in all these cases the basins of attraction are small, with the fraction of correctable errors either shrinking with network size or negligible to begin with. Thus these networks are not robust to noise. Very recently, a capacity of $C = \sim e^{\alpha \sqrt{N}}$ was realized in Hopfield networks with particular clique structures \cite{Hillar14robust,Fiete14}. As required by information-theoretic constraints, the memory states do not store arbitrary input patterns. The information rate ($\rho \equiv \ln(C)/N$) of these networks still vanishes with increasing $N$ and the networks are susceptible to adversarial patterns of error: Switching a majority of nodes within one clique (but still a vanishing fraction ($\sim \sqrt{N}/N$) of total nodes in the network) results in an unrecoverable error. 
 
The problem of robustly storing (or representing or transmitting) a large number of states in the presence of noise is also central to information theory. Shannon proved that it is possible to encode exponentially many states ($\sim e^{\alpha N}, 0<\alpha<1$) using codewords of length $N$ and correct a finite fraction of errors with the help of an optimal decoder \cite{Mackay04, Shannon48}. However, the complexity cost of the decoder (error-correction) is not taken into account in Shannon's theory. Thus, it remains an open question whether an associative network, which not only represents but also implements the decoding of its memory states through its own dynamics, should be able to achieve the same coding-theoretic bound. 

Here we combine the tradition of ACAM network theory with new constructions in coding theory to demonstrate a network with exponentially many stable states and basins that grow with network size so that errors in a finite fraction of all input neurons can be robustly corrected.

Note that portions of these results have been presented at conferences (R. Chaudhuri, I. Fiete Cosyne Abstracts, II-78, 2015; Soc. Neurosci. Program No. 94.05, 2015).

\section*{Results}

\subsection*{Linear error-correcting codes embedded in ACAM networks}

\begin{figure}
    \begin{center}
        \includegraphics[width=10 cm]{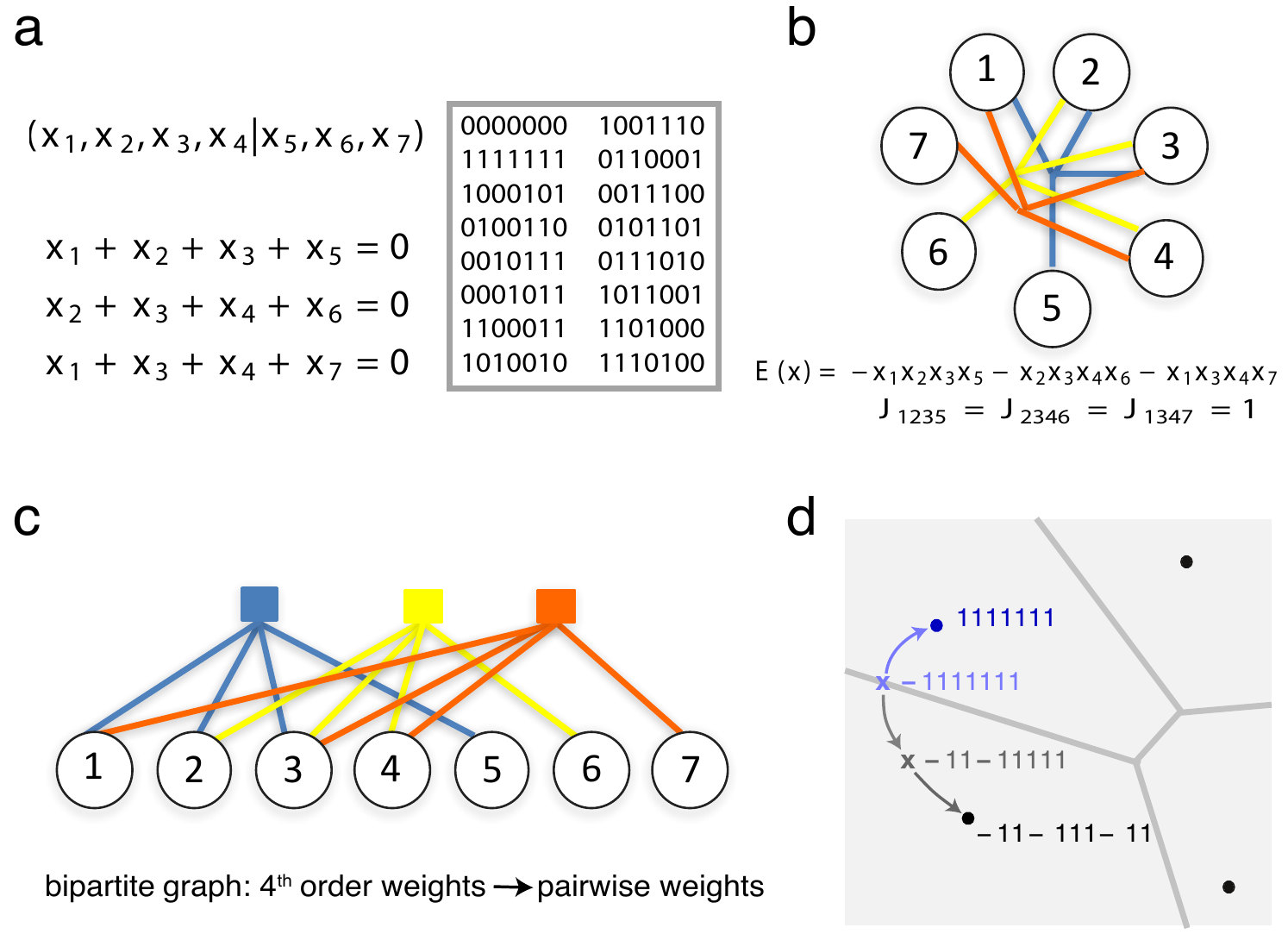}
    \end{center}
    \caption{{\bf Linking Hopfield networks to good error-correcting codes.} (a) Top left: The (7,4) Hamming code consists of 4 information-carrying bits and 3 additional bits for redundancy so that errors can be corrected. The 7-bit message must satisfy 3 constraint equations (bottom left); any solution to these equations is a permitted state or codeword (right). This code is capable of correcting all 1-bit errors (using a decoder that maps corrupted or noisy states to the nearest codeword) at a finite information rate of 4/7. (b) The Hamming codewords can be embedded as the stable states of a Hopfield network with 7 nodes, if the network is allowed to have 4th-order edges (to simplify the form of the energy function, states are mapped from $(0,1)$ to $(1,-1)$ for this Hamming implementation only). The edges simply enforce relationships between nodes (e.g. enforcing that $x_1, x_2, x_3, x_5$ share the same activation level), to implement the (4-point) constraint equations of (a). (c) The higher-order weights can be replaced by pairwise weights if the recurrent network is transformed into a bipartite graph with a hidden layer. Here each constraint node in the hidden layer (i.e., filled squares) is not a single neuron but a network that implements a parity operation on its inputs. A sample implementation of this is shown in Fig. 4. (d) Although the neural network has energy minima at every codeword, it can fail to correctly decode even 1-bit errors. The network dynamics follows any path that reduces energy, but that path can lead to a minimum that is farther away: In the example shown, a state ($-1111111$; lavender cross) one flip away from the closest codeword ($1111111$) can proceed along two trajectories that decrease energy, one to the nearest codeword (upwards) and the other to a codeword two flips away (downwards). Note that this problem is generic: starting from a non-codeword state in the energy landscape defined by a Hamming code of length N, N/2 directions lead to lower energy states but only 1 leads to the nearest codeword.
    }
\label{fig:hamming_hopfield}
\end{figure}

The codewords of linear error-correcting codes that use parity checks on binary variables can be stored as stable fixed points (equivalently, minima of the energy function) of a recurrent neural network. Consider embedding the $2^4$ codewords of the classic (7,4) Hamming code \cite{Hamming50} --- well-separated by 3 bits-flips from each other (Figure \ref{fig:hamming_hopfield}a) --- as stable fixed points in a network of 7 neurons. It is possible to do this using 4th order weights to enforce the relationships among subsets of four variables in the Hamming codewords (Figure \ref{fig:hamming_hopfield}b); the state of a neuron represents a corresponding bit in the codeword. By construction, codewords correspond to the energy minima of the network dynamics (Figure \ref{fig:hamming_hopfield}b). 

Alternatively, higher order ($K$th order, where $K>2$) relationships between multiple neurons can be enforced using only pairwise weights between a set of hidden layer neurons and $K$ visible neurons in a bipartite graph (Fig. \ref{fig:hamming_hopfield}c).

In either case, the network dynamics do not correctly decode states: starting one bit-flip from a stable state, the network might flip a second bit, a move that is also downhill in energy but further in (Hamming) distance from the original state. The dynamics then converge to a different codeword than the original (Figure \ref{fig:hamming_hopfield}d) -- a decoding error. Such errors are generic for Hamming codes. Starting one bit-flip from a stable state, over 50\% of possible bit flips lead to a state that has lower or equal energy, and all but one of these move the network away from the correct (nearest) stable state (see SI Section 4). The error can be attributed to a failure in {\em credit assignment}: the network is unable to identify the actual flipped bit. The network's failure should not be too surprising since in general decoding strong error-correcting codes is computationally hard \cite{Berlekamp78} and requires sophisticated but biologically implausible algorithms like belief propagation that do not map naturally onto simple neural network dynamics. 

\subsection*{An exponential-capacity robust ACAM}

We prove that it is possible to construct an ACAM network with stable states that grow exponentially in number with the size of the network, and that can be robustly corrected by the network's simple dynamics. Theoretical results establishing the number and robustness of these states are given in SI. Numerical simulation results on the number of stable states are shown in Figure 3a. Moreover, numerical simulations verify that the network dynamics corrects errors in a finite fraction of all the neurons in the network, meaning that the total number of correctable errors grows in proportion to network size, Figure 3b, up to a maximum corruption rate. Equivalently, given any error probability smaller than this rate, the network can correct all errors (with probability $\rightarrow 1$ as the network size goes to infinity). Typical of strong error-correcting codes, the probability of correct inference is step-like: errors smaller than a threshold size are entirely corrected, while those exceeding the threshold result in failure. Analogous results, up to small fluctuations, hold for stochastic dynamics in a Boltzmann Machine, SI Figure 2. 
  
The specific architecture of the network makes such performance possible. The network has a two-layer structure, consisting of one layer of $N$ input neurons and one layer of constraint neurons. The constraint neurons are organized into $N_C \sim N$ small sub-networks, which we call constraint nodes. Each constraint neuron in a constraint node makes connections with the same set of input neurons. Thus we first describe the connectivity between input neurons and constraint nodes, Figure 4a (top). The $i$th input neuron connects to $z^{(i)}$ constraint nodes, and the $j$th constraint node connects to $z_C^{(j)}$ inputs, and there are no within-layer connections between input neurons or across constraint nodes. The out-degrees $z, z_C$ can differ across input neurons and constraint nodes, and are drawn from narrow distributions with a fixed mean that does not scale with $N$; thus the network is truly sparse. The connectivity between layers is based on sparse expander graphs \cite{Hoory06, Lubotzky12}, mathematical objects with widespread applications in computer science.

\begin{figure}
    \begin{center}
        \includegraphics[width=12 cm]{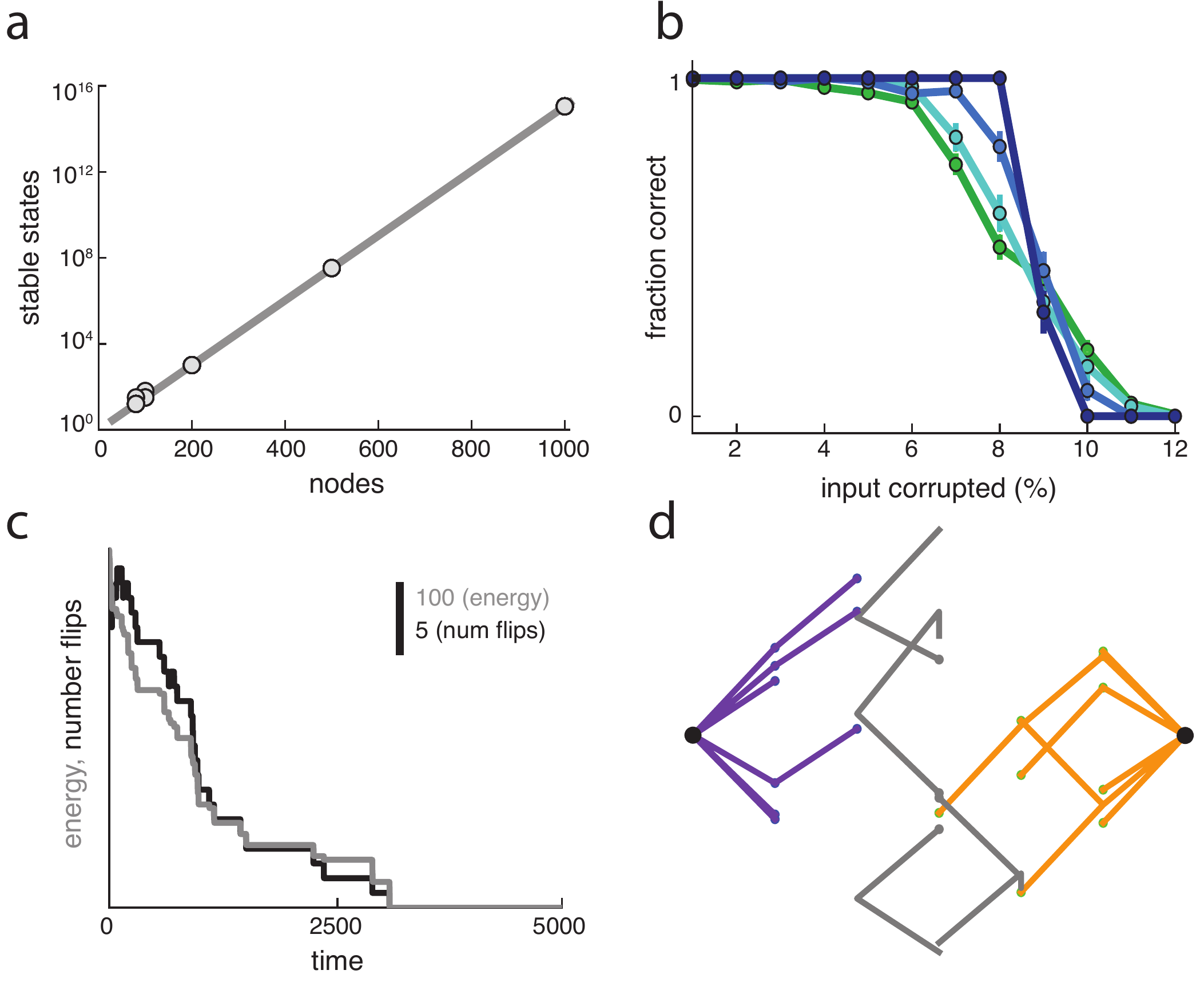}
    \end{center}
    \caption{{\bf ACAM with exponential capacity and robust error correction.}
(a) Network capacity scales exponentially with number of nodes. Gray line shows theoretical results for the number of stable states (Equation 2 with $r=1$ gives the number of stable states for a network with non-overlapping parity constraints; it provides a lower bound on the networks simulated here), while open black circles show the number of stable states in sample networks. Note that the scatter of points off the line for small N reflects occasional duplicate constraints in randomly-generated networks. For larger N these duplicates are vanishingly rare, and all points lie on the line.
(b) Fraction of times the network infers the correct state when a finite fraction of nodes (and thus a linearly growing number of total nodes with network size) are corrupted. Green, cyan, light blue, dark blue curves: $N=250, 500, 1000, 1500$ respectively. We choose the input node degrees to be between 5 and 10, and the constraint node degrees to be between 2 and 6 (see SI, S6 for more details on the parameters).
(c) Energy (gray) as a function of time in a simulated network of $N=500$ neurons with 4\% of all nodes initially corrupted, as the network relaxes to the closest stable state. Black: the number of node flips as the network state evolves. Note that energy always decreases monotonically, but the number of node flips need not. 
(d) Network state-space trajectories (projected onto 2D space) in a $N=18$ neuron network, starting from 6 different initial states (with 1-5 nodes corrupted) in the vicinity of one stable state (black dot, left), with an adjacent stable state (black dot, right). Initial states closest in Hamming distance to the original stable state flow to it (lines in varying shades of blue). Those that are not flow to other stable states (remaining lines).
}
\label{fig:performance}
\end{figure}

Once the out-degree distribution of input and constraint nodes is chosen, the connections between specific input neurons and constraint nodes is random. This procedure generically generates a sparse network with good expansion properties \cite{Sipser96, Luby01}. As described below, the expansion property is critical for allowing the network to robustly and correctly decode noisy states. 

Each constraint node is a subnetwork of neurons, all connected to the same subset of $z_C$ input neurons, Figure 4b. A constraint node can thus be viewed as a {\em glomerulus}. Thus, while the input-to-constraint node connectivity is random, at the level of individual constraint neurons connectivity within a node is correlated since all neurons must receive the same set of inputs -- the network is not fully random. Let each neuron in a constraint node be strongly activated by a different permitted configuration of the input neurons, and let the permitted configurations differ from one another by at least two flips -- comprising $\leq 2^{z_C-1}$ configurations (out of a set of $2^{z_C}$ total input configurations). The $\leq 2^{z_C-1}$ neurons in a constraint node are driven by their inputs and strongly inhibit each other (Figure 4c). The constraint node subnetwork can be small because $z_C$ does not scale with $N$. (In the numerical results of Figure 3, $z_C$ is distributed between $2$ and $6$.) 

The Lyapunov (generalized energy) function of the network is:
\begin{equation}
E(\xv, \hv) = - \left( \xv^T U \hv + \bv^T \hv + \frac{1}{2} \hv^T W \hv \right)
\label{eq:network_energy}
\end{equation}
where $\xv, \hv$ are the activations of the input neurons and the neurons across all constraint nodes, respectively; $\bv$ are biases in the constraint neurons; $U$ are the symmetric weights between input and constraint neurons; and $W$ are the lateral inhibitory interactions between neurons within the constraint nodes. We prove that ordinary Hopfield dynamics in this network, with appropriately chosen $W,U$, not only lowers the energy of the network states, but does so by mapping noisy states to the {\em nearest} (in Hamming distance) permitted state, so long as at most a small fraction of the input neurons are corrupted (SI Sections 6-8). Thus, the network is a good decoder of its own noisy or partial states, and can be called both associative and content-addressable. 

Permitted patterns form an exponentially large subset ($\sim 2^{\beta N}, 0<\beta<1$) of all possible binary states of length $N$, and are stable states of the network dynamics. The constraint nodes define the {\em permitted} states to be combinations, across nodes, of the set of preferred input configurations for each node. Constraint nodes correct corrupted input patterns to the nearest (in Hamming distance) permitted pattern. To understand how the  network dynamics achieves this functionality, first note that the input nodes change state much more slowly than the constraint nodes (see SI Section 8 for further details). Thus, we start by considering the activity of the constraint nodes for fixed input. The lack of coupling between constraint nodes means that each node is conditionally independent of the others given the inputs; thus for slowly-changing inputs the dynamics of each constraint node can be understood in isolation. When the input to a constraint node matches the preferred configuration of one of the constraint neurons, that neuron is maximally excited and silences the rest through strong inhibitory interactions within the node, Figure 4c. (It is possible to replace within-node inhibition with a common global inhibition across all neurons in all constraint nodes. This results in a slowing of the convergence dynamics, but not the overall quality of the computation). 

A single active winner neuron in a constraint node corresponds to a low-energy state that we call ``satisfied'' Figure 4c (left, green). If the input exactly matches none of the permitted (preferred) configurations, more than one constraint neuron with nearby preferred configurations will receive equal drive, Figure 4c (right, red). Global inhibition will not permit them all to be simultaneously active, but a pair of them can be; the network state then drifts between different pairs that are activated from among the equally driven subset of constraint neurons, and the node is in a higher energy, ``unsatisfied'' configuration (Figure 4c). 

We next consider the dynamics of the input neurons. An input neuron with more unsatisfied than satisfied adjoining constraint nodes will tend to flip under the Hopfield dynamics since doing so will make previously unsatisfied constraints satisfied, outnumbering the now-unsatisfied constraints, which lowers the total energy (Equation \ref{eq:network_energy}) of the network state (Figure 4d, top panel). Iterating this process provably (SI and Methods) drives the network state to the closest Hamming distance stable state. 

Note that, like most codes, these Hopfield networks are not perfect codes, meaning that the codewords and the surrounding points that map to them (i.e., the spheres in Fig. 1d) do not occupy the entire space of possible messages. Indeed, in high dimensions, the majority of the state space lies in between these spheres and the network has a large number of shallow local minima in the spaces between the coding spheres. 

\subsection*{Credit assignment and capacity with expander graph architecture} 

Why does the preceding network construction not fall prey to the credit-assignment errors exhibited by neural network implementations of Hamming codes? Our network can identify and correct errors by virtue of its sparse {\em expander graph} \cite{Sipser96,Luby01,Hoory06, Lubotzky12} architecture. 

A graph with good (vertex) expansion means that all small subsets of vertices in the graph are connected to relatively large numbers of vertices in their complements, Figure 4a (top). For instance, a subset of 4 vertices each with out-degree 3, that connects to 12 other vertices is maximally well connected and consistent with good expansion; the smallest possible set of neighbors for this subset would be 3 -- achieved if every vertex connected to the same set of 3 others -- an example of minimal expansion (Figure 4a, bottom).

Consider a bipartite graph with $N$ input vertices of degree $z$ and $N_C$ hidden vertices of degree $z_C$. $z$ and $z_C$ can vary by node, but $N\langle z \rangle = N_C\langle z_C\rangle$. Such a graph is a $(\gamma, (1-\epsilon))$ expander if every sufficiently small subset $S$ of input vertices (of size $|S|<\gamma N$, for some $\gamma<1$) has at least $(1-\epsilon)|\delta(S)|$ neighbors ($0 \leq \epsilon < 1$) among the hidden vertices, where $|\delta(S)| = \sum_{i \in S} z^{(i)}$ is the number of edges leaving the vertices in $S$ (if the graph is regular, meaning that $z$ and $z_C$ are constant, then $|\delta(S)| = z|S|$ where $|S|$ is the size of $S$) \cite{Sipser96,Luby01}.The deviation from maximal possible expansion is given by $\epsilon$, with $\epsilon \rightarrow 0$ corresponding to increasing expansion. We are interested in sparse expanders, where the number of connections each input unit makes scales very weakly or not at all with network size.

For sparse networks with high expansion, input neuron pairs typically share very few common constraint nodes (Fig. 4a and d). For a sufficiently small error rate on the inputs, many constraint nodes connect to only one corrupted input each. Because permitted states at each constraint node are separated by a distance of at least two input flips, many of these constraint nodes are unsatisfied, and the collection of unsatisfied constraint nodes can correctly determine which input neuron should flip. This property of expander graphs provably allows for simple decoding of error-correcting codes on graphs if the constraint nodes impose parity checks on their inputs  \cite{Sipser96}; we establish that simple Hopfield dynamics in neural networks implementing more general (non-parity) constraints can achieve similar decoding performance (SI, Sections 6-8).  

By contrast, if two or more corrupted input neurons project to the same constraint node, the resulting state may again be permitted and the constraint satisfied (Fig. 4d, bottom panel). The corrupt input nodes are now deprived of an unsatisfied constraint that should drive a flip, leading to a potential failure in credit assignment. These failures are far more likely if the graph is not a good expander. 

Returning to why the neural network Hamming code failed to properly error-correct, note that a variable in a Hamming code has about 50\% probability of connecting to each constraint. Thus a single error bit makes about half the constraints unsatisfied. Flipping a randomly-chosen second bit will, on average, change the state of half the constraints, leaving the mean number of satisfied constraints unchanged. Thus, with at least 50\% probability, a random bit flip leads to a state with energy less than or equal to the single error state. Note that a similar argument should apply to any code on a dense graph (see SI Section 4 for more details). 

If the $j$th constraint node has $2^{z^{(j)}_C-r^{(j)}}$ permitted configurations (where $r^{(j)} \geq 1$ is some real number), the average total number of energy minima across constraints equals
\begin{equation}
2^{-\mean{r}N_C} 2^N = 2^{\left(1-\mean{r}\frac{\mean{z}}{\mean{z_C}}\right) N}=2^{\left(1-\mean{r}\hat{z}\right) N},
\end{equation}
where we have defined $\hat{z} = \frac{\mean{z}}{\mean{z_C}}$. Each constraint node has at most $M=2^{(z^{max}_C-1)}$ neurons and so the total network size is $N_{net} \leq N + MN_C$. Therefore the number of minima is
\begin{equation}
N_{\text{states}} \geq 2^{\alpha N_{net}} \quad \text{where} \quad \alpha=\frac{\left(1-\mean{r}\hat{z}\right)}{\left(1+M\hat{z}\right)}.
\label{eq:capacity}
\end{equation}

The number of minimum energy states is exponential in network size because $\alpha$ is independent of $N, N_{net}$.

Given a network with $N$ variables and $N_C$ constraint nodes that is a $(\gamma, (1-\epsilon))$ expander, network dynamics can correct the following number of errors in the input neurons:
\begin{equation}
N_\text{errors}\geq \beta N \quad \text{where} \quad \beta=\gamma (1-2 \epsilon).
\end{equation}
Thus, the number of correctable errors grows with network size, with proportionality constant $\beta$ that depends on network properties but not on network size. In SI Figure 1 we show numerical estimates of expansion for the graphs we use to generate the results in Figure 3.

Briefly clamping the state of the input neurons before allowing the network to run determines the state of the constraint nodes, thus initial errors in the constraint node states do not affect network performance. Note that these results require an expansion coefficient $\epsilon > 1/4$, (in SI, Section 5 we summarize results showing that this expansion is generically achieved in random bipartite graphs), permitted input configurations for each constraint that differ in the states of at least two inputs neurons, and no noise in the update dynamics (meaning that each node always switches when it is energetically favorable to do so). In SI Sections 9 and 10, we extend these results to weaker constraints and to noisy (Boltzmann) dynamics.

\begin{figure}
    \begin{center}
        \includegraphics[width=9 cm]{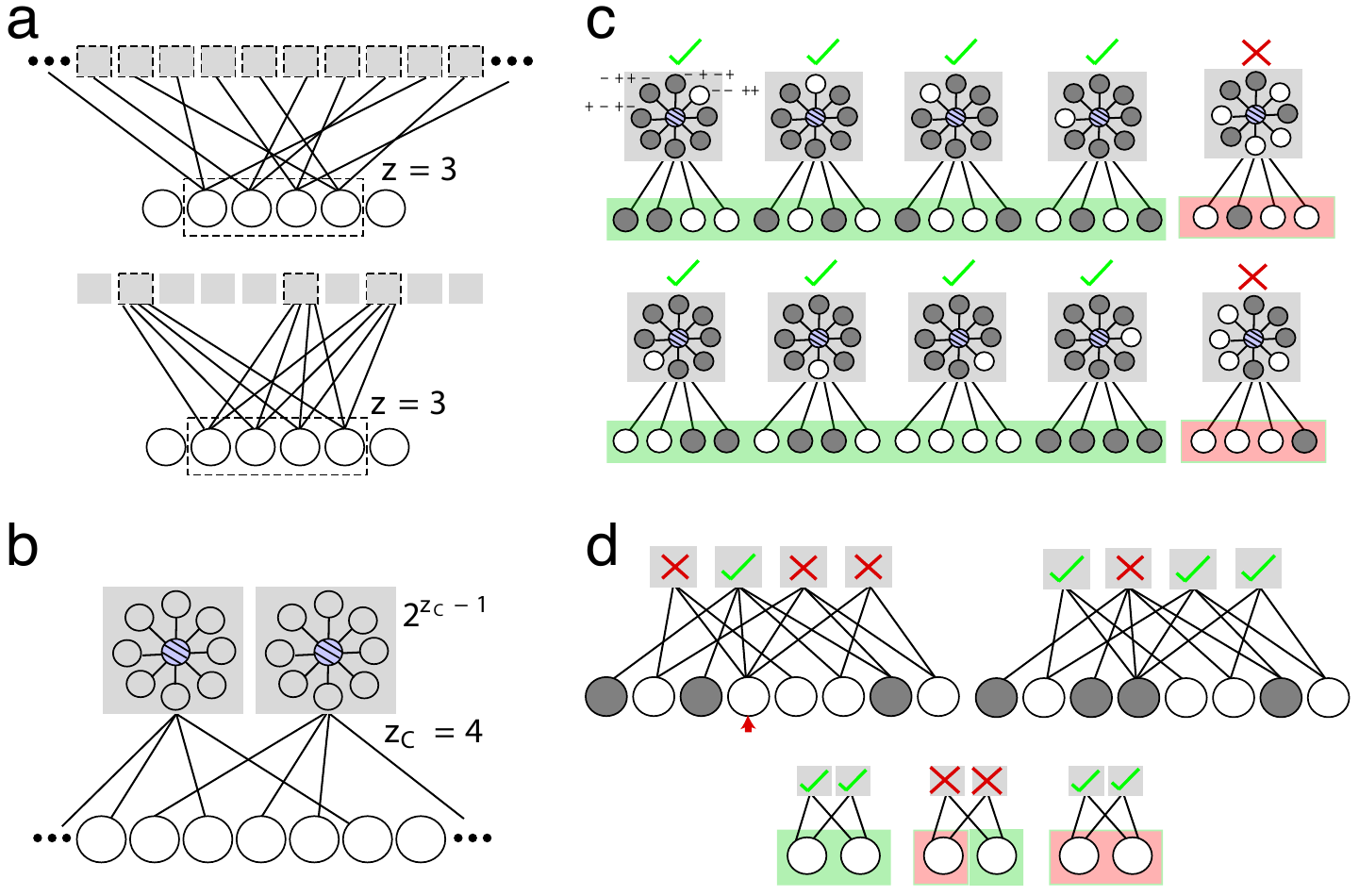}
    \end{center}
    \caption{{\bf Architecture of the robust exponential ACAM.} 
(a) Two bipartite networks, each with $N$ input neurons (bottom layer, circles) and $N_C$ constraint nodes (top layer, squares). The out-degree for the input nodes is set to $z= 3$ in both networks. The top network exhibits excellent expansion: subsets of input nodes (one subset highlighted by dashed line) have many neighbors in the constraint layer. Equivalently, input neurons share relatively few common constraints. The network at bottom exhibits poor expansion: the same-sized subset of inputs has many fewer neighbors. 
(b) Each constraint node is a subnetwork of several sub-nodes or neurons. The number of neurons in the constraint node is 
$2^{z_C-1}$, where $z_C$ is the in-degree of the constraint node (a value that does not scale up with network size). Neurons in a constraint node receive global inhibition (hashed blue node). 
(c) Response of one constraint node to different configurations of its inputs. Each constraint neuron prefers one input configuration, determined by the weights from the inputs to that neuron (shown above a few neurons as a pattern of $+$'s and $-$'s): a positive (negative) weight from neuron $i$ causes the neuron to prefer input $i$ to be ``on'' (``off''); on (off) states are shown in white (gray). Here we have depicted all even-parity input states as the set of preferred states (green background). In a preferred input state, only one constraint neuron is strongly driven  (white), corresponding to a low-energy or ``satisfied'' constraint (green check). For a non-preferred input (red background), multiple constraint neurons with preferred states that closely resemble the input are equally driven (white). Global inhibition forces no more than two of these constraint neurons to be active, and the constraint node wanders across configurations where pairs of these neurons are active. This is a higher-energy, unsatisfied constraint state (red cross). 
(d) Top left: network state with satisfied and unsatisfied constraints. An input attached to more unsatisfied than satisfied constraint nodes (red arrow, bottom) lowers the energy of the network when it flips, by flipping the status of all its constraint nodes (top right). Iteration of this process drives the network to the nearest Hamming-distance stable state. Bottom: Each constraint node is a weak check on its inputs: starting from a pair of satisfied constraints (left), corrupting a single input violates the constraints (center), but corrupting a second input (right) erases the signature of input error in the constraints. The problem of non-expansion: if two constraints have overlapping input states, they cannot help identify the source of input error (bottom center).
}
\label{fig:net_arch}
\end{figure}

\subsection*{Self-organization to exponential capacity}
We next show that a network with initially specified connectivity but unspecified weights can self-organize to have exponentially many well-separated minima. This self-organization can be performed by a simple one-shot learning rule that depends on the coactivation of pairs of input and constraint neurons. We start with a bipartite architecture consisting of $N$ input neurons in one layer and $N_C$ constraint nodes of $M$ neurons each in the other layer. An input neuron sends connections to $z$ randomly chosen constraint nodes, contacting all the neurons in the constraint node with weak non-specific connection strengths. Conversely, each neuron in a constraint node receives $z_C$ connections from input neurons, with all neurons in one constraint node connected to the same subset of inputs. We choose the number of neurons in a constraint node $M \geq 2^{z_C-1}$. Note that the degrees $z$ and $z_C$ do not need to be fixed (but we consider them fixed for simplicity).

\begin{figure}
    \begin{center}
        \includegraphics[width=8 cm]{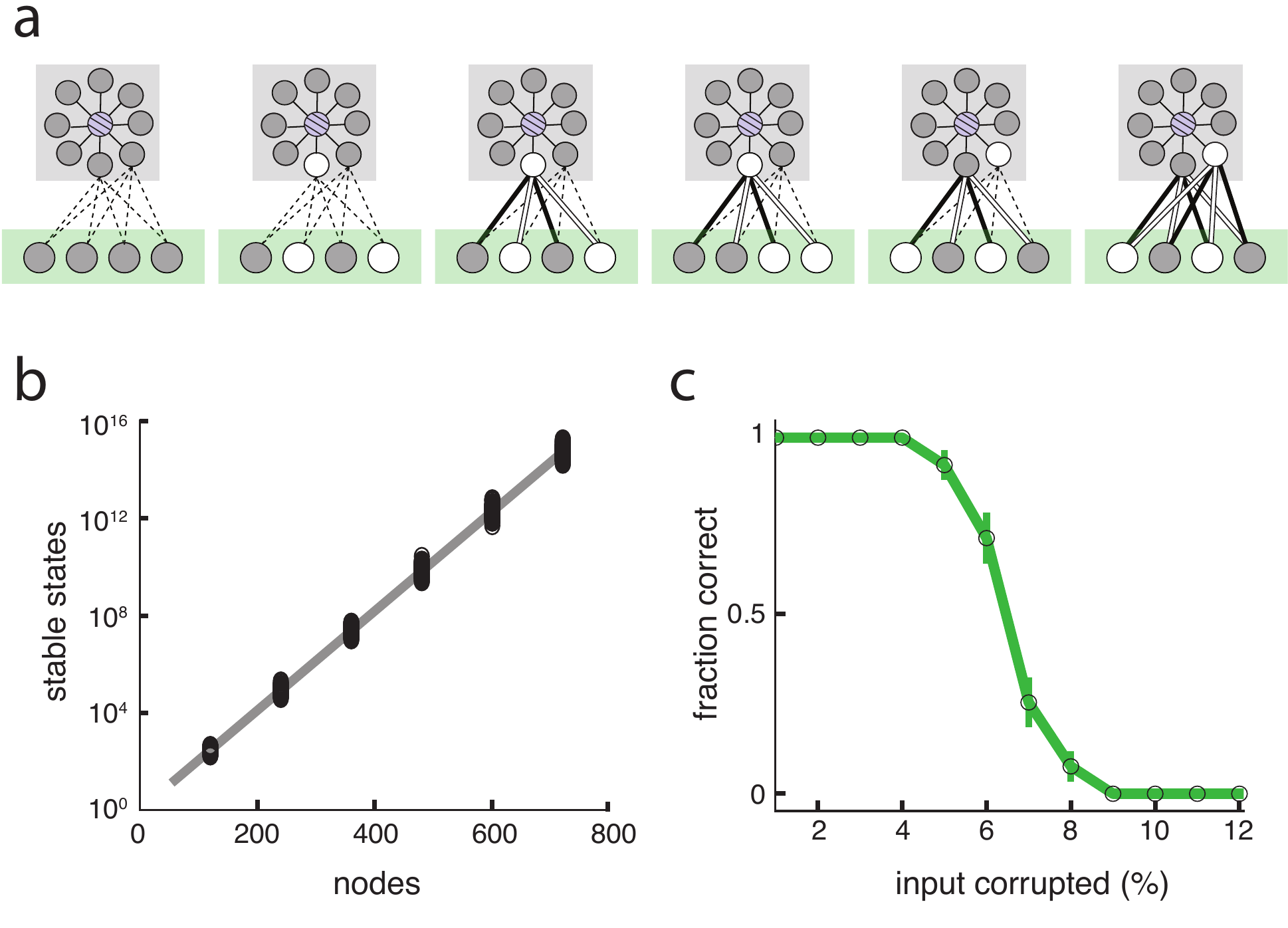}
    \end{center}
    \caption{{\bf Input-driven self-organization of weights in a network with random connectivity.} 
(a) Schematic of learning at a single constraint node. Initially the constraint node receives weak, non-specific projections from a subset of input nodes (first panel). Learning then proceeds by pairing random input patterns with sparse random activation of constraint neurons. If the input to a constraint node is not close to a previously learned pattern and a constraint neuron is active, the active constraint neuron learns connection strengths that prefer this pattern (second and third panels; fifth and sixth panels). If the constraint node receives an input that is close to a previously learned pattern, then the constraint neuron corresponding to this pattern activates and suppresses the activation of other constraint neurons (fourth panel).
(b) Average number of fixed points for learned network. 
(c) Error-correction performance for learned network with N=480.}
\label{fig:learning}
\end{figure}

During learning, we pair random sparse activation of constraint neurons with random patterns in the input neurons. When a constraint neuron is active, we set the weights of its connections to active input neurons to +1 and those to inactive input neurons to -1, in a Hebbian-like one-shot modification (Fig. \ref{fig:learning}a). We also set a background input to the constraint neurons so that all constraint neurons receive the same average input over time, regardless of their (learned) preferred input configurations. Once a synapse strength has been learned, it is no longer updated.

When a constraint neuron receives an input that matches or is close to (one input neuron flip away from) a previously seen and thus learned input, it will become active, suppressing the other constraint neurons in its node and preventing further learning. Consequently each constraint node learns to prefer a subset of possible input states that it sees, so long as they differ in at least two entries. This procedure is equivalent to each constraint node choosing a random subset of the presented input states, with a minimum Hamming distance of 2 between them (see SI, Section 11 for more details). As a whole, if the piece of an input pattern that one constraint sees is called a fragment, the network learns to prefer combinations of fragments from across all input patterns. Note that initially this process is greedy, with the network learning each presented fragment. However, later in the learning, presented fragments start to conflict with previously learned fragments (i.e., they differ in fewer than two entries), and the network ignores them. 

As shown in Fig. \ref{fig:learning}b and c, this yields a network with a capacity (number of robust memory states) that grows exponentially with the number of neurons and the ability to recover from errors in a finite fraction of all neurons.

\subsection*{Robust retrieval of labels for noisy input patterns}
As we describe in the Introduction (also see Discussion), it is impossible for a network with $N$ neurons to store more than $O(N)$ arbitrary patterns as memories. The patterns stored in our exponential capacity network are not arbitrary: they are determined by a large number of sparse constraints. Given these restrictions on the structure of patterns that can be stored at high capacity, how might these networks (or any such high capacity networks) be used?

One possibility is that the networks are used as content-addressable memories for a class of inputs with a particular structure, and we consider this possibility further in the Discussion.

\begin{figure}
    \begin{center}
        \includegraphics[width=12 cm]{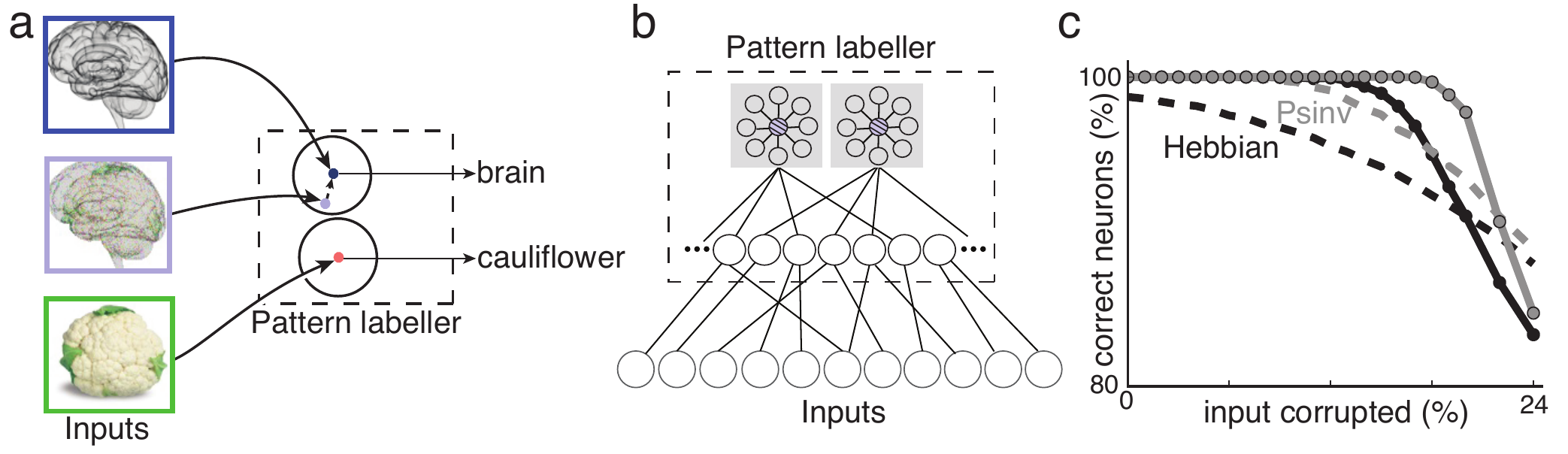}
    \end{center}
    \caption{{\bf Robust labelling of input patterns.} 
(a) Schematic of mapping input patterns from an $M=e^N$ dimensional input space to exponentially-many well-separated storage patterns in an $N$-dimensional network.
(b) Structure of multi-layer network that carries out the mapping in (a). 
(c) Average number of errors in the input to the labelling network after feedforward mapping but before error correction in the labelling network (dashed line) and average number of errors after the network is allowed to perform error correction (solid line), both shown as a function of the number of flips in the M-dimensional input space. Here $M=10,000$, $N=480$ and number of stored patterns $= 2000$. Black and gray traces show Hebbian and pseudoinverse constructions respectively.}
\label{fig:pattern_label}
\end{figure}

Alternatively, the pre-structured repository of robust network memory states can serve as a neural pattern labeler (or locality-sensitive hash function), in which distributed input patterns are assigned abstract indices corresponding to the memory states, Fig. \ref{fig:pattern_label}a,b. For example, some general theories of the hippocampus see it as assigning an index or hash value to sparse distributed patterns of cortical input \cite{Valiant12}. However, for this to be possible with the relative sizes of cortex and the much-smaller hippocampus (the cortex in rats contains $10-100$ times more neurons than the hippocampus --- $\sim 5\times 10^7$ in cortex versus $\sim 10^6$ in hippocampus \cite{West91,Herculano09}, while in humans the factor is $\sim 1000$ --- $\sim 10^{10}$ versus $\sim 10^7$ neurons\cite{West90, Pakkenberg97,Herculano09}) requires a high-capacity indexing scheme like the one we describe below. 

Consider a set of $O(M)$ input patterns in a network with $M$ neurons, where $M$ is possibly very large. For example, cortical representations are typically sparse, and thus these input patterns could be a set of sparse cortical representations distributed across a large number of neurons. Or they might consist of non-sparse activity states lying on some low-dimensional subspace of the space of all patterns. We wish to map these input patterns to the (exponentially-many) stable states in a much smaller memory network. If the memory state has $N$ neurons, then $M$ can be $O(e^N)$. 

An appropriate feedforward mapping from the $M$-dimensional input network to the $N$-dimensional memory network can be constructed using a simple Hebbian or correlational learning rule that updates synaptic strengths using the product of the desired input and output states\cite{Hopfield82}, numerical simulations in Fig. \ref{fig:pattern_label}c and see SI S12 for proofs. Such a mechanism would require the memory network to be able to spontaneously move to new states to generate new labels (reminiscent of the observation that spontaneous plateau potentials in CA1 determine new place cells\cite{Bittner15}). Alternatively, if the set of input patterns are known ahead of time, the synapses for the feedforward mapping can be defined using a pseudoinverse construction; this produces a more noise-tolerant mapping than the Hebbian learning rule \cite{Kanter87}, as seen by contrasting the black and gray traces in Fig. \ref{fig:pattern_label}c (and see SI S12). Either of these mappings involves $O(MN)$ total synapses (with $N<<M$); moreover, if the memory network is spatially-localized compared to the input network, then this scheme conserves wiring length compared to global connections in the input network. 

If the input patterns are well-separated (as will be true for any generic set of $M$ patterns, see SI S12), then both the Hebbian and pseudoinverse mappings preserve the local neighborhood structure, in the sense that locally perturbed versions of an input pattern map to the local neighborhood of the corresponding label (SI S12). Thus, given a noisy input pattern within a neighborhood of the original, the state in the memory network flows to its stable state, retrieving the correct label (schematic in Figure \ref{fig:pattern_label}a, numerical results in Fig. \ref{fig:pattern_label}c and proofs in SI S12). The noise tolerance is linear in the dimension of the input space, meaning that it is possible for some finite fraction of the very large number of input neurons to be wrong. Consequently, the number of input errors can be much larger than the number of neurons in the memory network. 

Thus, the memory network uses its exponential capacity to robustly index a much large number of input patterns. Note that this mapping is one-way: the labels can be recovered from the (possibly perturbed) input patterns but, in accordance with information-theoretic bounds \cite{Abu85, Gardner88}, the labels cannot be used to recover the input pattern. Instead, the compressed labels generated and robustly retrieved by the exponential-capacity network in response to noisy or incomplete input patterns can then be used to drive downstream associations and actions using a much smaller number of synapses. Such a network could also be used for template matching, classification, locality-sensitive hashing and nearest neighbor computations (indeed, locality-sensitive hashing can be used to compute fast approximate nearest-neighbors\cite{Andoni06}), and for any computation where sparse patterns on a large space need to be compressed into dense patterns on a smaller space. 

\section*{Discussion}
Unlike current theoretical neural architectures, which either show a small number of stable states (i.e. sub-exponential in network size) or weak error correction (i.e. the number of correctable errors is a vanishing fraction of network size), we demonstrate that neural networks with simple Hopfield dynamics can combine exponential capacity with robust error-correction. In the networks we construct, each constraint only weakly determines the state of the small number of nodes it is connected to: it restricts the states to a large subset of all possible states. However, the decorrelated structure of network connectivity, due to its expander graph architecture, means that these constraints are not strongly overlapping, and they can specify and correct patterns of errors on the inputs. In short, the network combines a large number of sparse, weak constraints near-optimally to produce systems with high capacity and robust error correction. 

General arguments show that recurrent neural networks with $N$ neurons cannot store more than $O(N)$ arbitrary patterns as memories \cite{Abu85, Gardner88}. A rough explanation of this limit is that fully-connected networks of $N$ neurons have $N^2$ synapses, which are the free parameters available for storing information. If each synapse has a finite dynamic range, the whole network contains $O(N^2)$ bits of information (with a proportionality constant determined by the number of states at each synapse). An arbitrary binary pattern over $N$ neurons contains $N$ bits of information, therefore the network cannot store more than $O(N)$ such patterns. The networks we construct do not contradict these results even though the number of robust stable states is exponential: The patterns represented are not arbitrary. In other words, as with any super-linear capacity results \cite{Hillar14robust, Fiete14}, the networks we construct exhibit robust high-capacity storage for patterns with appropriate structure rather than for random patterns. 

Is it possible to circumvent bounds on the storage of fully random patterns through an alternate scheme, in which exponentially many arbitrary patterns are mapped to these robust memory states? Encoders in communications theory do just this, mapping arbitrary inputs to well-separated states before transmission through a noisy channel. From a neural network perspective, the feedforward map can be viewed as a recurrent network with input and hidden units and asymmetric weights, so again we know from capacity results on non-symmetric weights \cite{Gardner88} that it should not be possible. Mapping exponentially many arbitrary patterns to these structured memory states in a retrievable way would require specifying exponentially many pairings between inputs and structured memory states, and thus in general, equally many synapses. One way to obtain that many synapses would be to have exponentially many input neurons, but then the overall network would not possess exponential capacity for arbitrary many patterns as a function of network size. Note that this is the case for the neural pattern labeler we discuss above, where there are exponentially-many input neurons that are not part of the memory network, and thus the number of synapses required is smaller than the total squared network size. Put another way, while random projections of $\sim e^M$ points into an $M$ dimensional space preserve the relative distances and neighborhoods of these points \cite{Johnson84}, which means that exponentially well-separated points can remain well-separated in a logarithmically shrunken dimensional space through a simple linear projection, such projections will not generally result in a specific expander structure of the embedding to allow for error correction by simple neural network dynamics. Arguably, however, brains are not built to store random patterns, and natural inputs that are stored well are not random. 

As shown in Fig. \ref{fig:pattern_label}, these networks can be used to to generate robust labels for arbitrary input patterns in a high-dimensional space. 
Another possible use for these networks is as content-addressable memories for inputs with appropriate structure. These inputs must be well-described as a product of a large number of sparse, weak, decorrelated constraints. For instance, natural images are generated from a number of latent causes or sources in the world, each imposing constraints on a sparse subset of the pixelated retinal data we receive. Alternatively, it must be possible to transform the set of inputs to have appropriate structure. For our networks, this would correspond to decorrelating structure in the lower moments of the data \cite{Barlow61,Olshausen96, Vinje00} while preserving structure in higher moments. It is still an open question what kinds of stimuli may either be naturally described within this framework or easily transformed to have the appropriate structure, but recent results suggest that the structure of natural images can be captured by the minima of Hopfield networks \cite{Hillar14images}.

As observed by Sourlas \cite{Sourlas89,Sourlas01}, Hopfield networks and spin glasses can be viewed as error-correcting codes, with the stable states corresponding to codewords and the network dynamics corresponding to the decoding process. Moreover, it is possible to embed the codewords of general linear codes into the stable states of Hopfield networks with hidden nodes or higher-order connections (illustrated by our embedding of the Hamming code and shown in SI, Section 3) -- thus, there can be exponentially many well-separated stable fixed points. However, decoding noisy inputs is a hard problem for high-capacity codes (decoding general linear codes is NP-hard \cite{Berlekamp78}), requiring complex inference algorithms. These algorithms do not map naturally onto Hopfield dynamics, and as a result (as seen for the Hamming code) codewords cannot in general be correctly decoded by the internal dynamics of the neural network.

We leveraged here recent developments in high-dimensional graph theory and coding theory, on the construction of high-capacity low-density parity check codes (expander codes) that admit decoding by simple greedy algorithms \cite{Sipser96, Luby01}\footnote{However, a larger number of errors are correctable, or equivalently, the correctable basins of the codewords are larger with more complex techniques like belief propagation.}, to show that ACAM networks with quasi-random connectivity can implement expander codes. If only pairwise connections are allowed then such a network requires hidden nodes. If higher-order connections are allowed then we show that sparse random ACAM networks (without hidden nodes) are isomorphic to expander codes (SI, Section 3) and generically have exponential capacity. Thus ACAM networks can have capacity and robustness performance comparable to state-of-the-art codes in communications theory, moreover with the decoder built into the dynamics. 

For the case of Boltzmann dynamics, the network we construct is a Restricted Boltzmann Machine \cite{Smolensky86, Hinton02} with constrained outdegree and inhibition (for sparsity) in the hidden layer. The network can be considered a product of experts \cite{Hinton02}, where each constraint node is an expert, and different neurons within a constraint node compete to enforce a particular configuration on their shared inputs. A constraint node as a whole constrains a small portion of the probability distribution. 

Several recent ACAM network models are based on sparse bipartite graphs with stored states that occupy a linear subspace of all possible states \cite{Karbasi13, Salavati14}. These networks exhibit exponential capacity for structured patterns, but either lack robust error correction \cite{Salavati14} or rely on complex non-neural dynamics with multiple stages \cite{Karbasi13}. 

The network of Hillar \& Tran \cite{Hillar14robust} is notable in that it is close to exponential capacity, has large basins of attraction, shows very rapid convergence and is easy to construct. This network can also be understood in terms of a sparse constraint structure, with participation in a constraint determined by membership in the clique of an associated graph. These recent results suggest that sparse constraint structures can be leveraged in multiple ways to construct high capacity neural networks.

Expander graph neural networks leverage many weak constraints to provide near-optimal performance and exploit a property (i.e., expansion) that is rare in low dimensions but generic in high dimensions (and hence can be generated stochastically and without fine-tuning); both are common tropes in modern computer science and machine learning\cite{Candes06, Donoho06,Kuncheva04, Schapire90}, and expander graphs have found widespread recent use in designing algorithms\cite{Hoory06,Sipser96}, including in solving challenging memory problems\cite{Larsen16}. Intriguingly, large sparse random networks are generically expander graphs \cite{Hoory06, Lubotzky12}, making such architectures promising for neural computation, where networks are large and sparse. The networks that we construct may thus provide broader insight into the computational capabilities of sparse, high-dimensional networks in the brain\cite{Litwin17}. 

The network we construct has $N$ input/variable neurons and $N_C \sim N$ constraint nodes, each containing $2^{z_c-1}$ neurons. Thus this architecture has many more constraint cells than input cells. As a result, one key prediction of our model is that most cells are sparsely active, and their role is to impose constraints on network representations (this sparse strategy is feasible only in high dimensions/large networks, where exponential capacity on even a small fraction of the network can yield enough gains to outperform classical coding strategies). These constraint cells are only transiently active while they filter out irrelevant features, while cells that carry the actual representation will respond stably for a given state. Consequently, representations are predicted to contain a dense stable core, with many other neurons that are transiently active. This is reminiscent of observations in place cell population imaging \cite{Ziv13} and the sparse distribution of population activities across hippocampus and cortex \cite{Buzsaki14, Rich14}. Neurons in each constraint node receive connections from the same subset of input neurons; this connection pattern resembles glomeruli, such as seen in the olfactory bulb and cerebellum \cite{Shepherd10}. On the other hand, the particular set of inputs to each constraint node are decorrelated or random, as required for good error correction. Thus the network architecture predicts clustered connections at the level of constraint neurons and, on the other hand, that input neurons should not share many of these constraint clusters in common. Finally, these architectures predict that representations in the input neurons should have relatively decorrelated second-order (i.e. pairwise) statistics, but should contain structure in higher-order moments that the network exploits for error correction. 

In summary, we bridge neural network architectures and recent constructions in coding theory to construct a robust high-capacity neural memory system, and illustrate how sparse constraint structures with glomerular organization might provide a powerful framework for computation in large networks of neurons.

\section*{Methods}
\subsection*{Hopfield networks and Boltzmann machines}
We consider networks of $N$ binary nodes (the neurons). At a given time, $t$, each neuron has state $\vect{x}_i^t$=$0$ or $1$, corresponding to the neuron being inactive or active respectively. The network is defined by an $N$-dimensional vector of biases, $\vect{b}$, and an $N\times N$ symmetric weight matrix $W$. Here $\vect{b}_i$ is the bias (or background input) for the $i$th neuron (equivalently, the negative of the activation threshold), and $W_{ij}$ is the interaction strength between neurons $i$ and $j$ (set to 0 when $i=j$).

Neurons update their states asychronously according to the following rule:
\begin{equation}
\vect{x}_i^{t+1}=
\begin{cases}
1 & \text{if } \sum_j W_{ij}\vect{x}_j^t + b_i>0\\
0 & \text{if} \sum_j W_{ij}\vect{x}_j^t +b_i < 0 \\
\text{Bern}(0.5) & \text{if} \sum_j W_{ij}\vect{x}_j^t + b_i = 0
\end{cases}
\end{equation}
Here Bern(0.5) represents a random variable that takes values $0$ and $1$ with equal probability.

Hopfield networks can also be represented by an energy function, defined as
\begin{equation}
E(\vect{x}|\vect{\theta},W)=-\frac{1}{2}\sum_{i\neq j} W_{ij}\vect{x}_i\vect{x}_j - \sum_i \vect{b}_i \vect{x}_i
\label{eq:hopf_energy}
\end{equation}
The dynamical rule is then to change the state of a neuron if doing so decreases the energy (and to change the state with $50\%$ probability if doing so leaves the energy unchanged).

We also consider Boltzmann machines, which are similar to Hopfield networks but have probabilistic update rules.

\begin{align}
\vect{x}_i^{t+1}&=\text{Bern}(p) \nonumber \\
\text{where } p&=\frac{1}{1+e^{-\beta(\sum_j W_{ij}\vect{x}_j^t - b_i)}}
\end{align}

The probability of a state $\vect{x}$ in a Boltzmann machine is $p(\vect{x}) \propto e^{-\beta E(\vect{x})}$, where $E(\vect{x})$ is defined as in Eq. \ref{eq:hopf_energy} and $\beta$ is a scaling constant (often called inverse temperature). Note that the Hopfield network is the $\beta \rightarrow \infty$ limit of a Boltzmann machine.

\subsection*{Hopfield network implementation of Hamming code}
The (7, 4) Hamming code can be defined on sets of 7 binary variables by the equations 
\begin{align}
x_1+x_2+x_3+x_5&=0 \nonumber \\
x_2+x_3+x_4+x_6&=0 \nonumber \\
x_1+x_3+x_4+x_7&=0, 
\end{align}
where all the sums are taken mod 2. 

This is equivalent to a 4th-order Hopfield network with the energy function 
\begin{equation}
E=-s_1s_2s_3s_5-s_2s_3s_4s_6-s_1s_3s_4s_7
\end{equation}
Only here, for the sake of a tidy form for the energy function, we let the variables $s_i$ take values in $\{-1, +1\}$ rather than in $\{0, +1\}$ (it is straightforward to reexpress the energy function with $\{0, +1\}$ states, but the expression is less tidy). In the rest of this text, nodes take the values $\{0, +1\}$. 

\subsection*{Hopfield network expander codes: construction}
We consider a network with two layers: an input layer containing $N$ nodes that determine the states or memories that will be stored and corrected, and $N_C$ constraint modules that determine the permitted states of variables (see Figure 3). These constraint nodes are themselves small sub-networks of nodes (see below). 

Each input neuron participates in $z$ constraints (i.e., is connected to $z$ constraint nodes), and each constraint node receives input from $z_C$ input neurons. Thus, the input neurons have degree $z$ and the constraint nodes have degree $z_C$; consequently $N\langle z \rangle = N_C\langle z_C\rangle$ (i.e., the number of edges leaving the input neurons equals the number of edges entering the constraint nodes). $z$ and $z_C$ are small (for Figure 3, $5\leq z \leq 10$ and $2\leq z_C \leq 6$, while for Figure 5, $z=5$ and $z_C=12$) and do not grow with the size of the network (i.e., with $N$ and $N_C$). Thus these networks are sparse.

The connections between the $N$ input neurons and the $N_C$ constraint nodes are chosen to be random, subject to the constraints on the degrees. Thus, with probability asymptotically approaching $1$ as $N$ increases, these networks are $(\gamma, (1-\epsilon))$ expanders with $\epsilon<1/4$ (see SI S5 and \cite{Sipser96, Luby01}).

There are $z_C$ variables connected to a given constraint node and these could take any of $2^{z_C}$ possible states. The constraint nodes restrict this range, so that a subset of these states have low energy (and are thus preferred by the network). Each constraint node is actually a network of neurons with Hopfield dynamics and, while there are multiple possible ways to construct constraint nodes, in the construction that we show, each neuron in the constraint node prefers one possible configuration of the input neurons. For simplicity we choose these preferred states to be the parity states of the inputs (i.e. states where the sum of the inputs modulo 2 is 0), but note that the set of preferred states can be chosen quite generally (SI, Sections 6 and 9).

If a neuron in a constraint node prefers a particular state of its input neurons, then it has connection weights of +1 with the input neurons that are active in that state and weights of -1 with those that are inactive. Thus, a constraint neuron that prefers all of its input neurons to be on will receive an input of $z_C$ when its input is in the preferred state, and a constraint neuron that prefers only $k$ of these nodes to be active will receive input of $k$. To ensure that all constraint neuron receive the same amount of input in their preferred state, we also add biases of $z_C-k$ to each constraint node, where $k$ is the number of non-zero variables in its preferred state. Finally, to ensure that multiple constraint neurons are not simultaneously active during a preferred input for one sub-node, we add inhibitory connections of strength $z_C-1$ between all the sub-nodes in a constraint node. As a consequence of this strong inhibition, each constraint node has competitive dynamics: in the lowest energy state the input neurons are in a preferred configuration, the sub-node in the constraint node corresponding to this configuration is active, and all other sub-nodes are suppressed.

The network is in a stable or minimum energy state when all of the constraints are satisfied. As shown in Eqs. 2 and 3 in the main text, if the $j$th constraint node has $2^{z^{(j)}_C-r^{(j)}}$ permitted configurations (where $r^{(j)} \geq 1$ is some real number), then the network has an exponential number of minimum energy states.

\subsection*{Hopfield network expander codes: error correction}
The proof that these networks can correct a constant fraction of errors is based on Sipser \& Spielman (1996) \cite{Sipser96} and Luby et al. (2001) \cite{Luby01} with slight generalization. We leave the details of the proof to the Supplementary Information, and sketch the main steps here.

We consider a set of corrupted input neurons, $S$, with size $|S|<\gamma N$. Since all subsets of size $<\gamma N$ expand, we can show that $S$ is connected to a comparatively large set of constraints, which we call $T$. In order for $|T|$ to be large, it must contain a large number of constraint nodes that are only connected to one neuron in $S$ (i.e. the neurons in $S$ do not share many constraint nodes in common), which we call non-shared. Constraint nodes can detect one error, so the non-shared nodes will be unsatisfied. If the fraction of non-shared constraint nodes is high, then at least one neuron in $S$ must be connected to more unsatisfied than satisfied constraint nodes, and it is energetically favorable for it to change its state. Thus, there is always an input neuron that is driven to change its state, reducing the number of unsatisfied constraints.

However, this does not exclude the possibility that the wrong input neuron changes its state. The remaining step of the proof is to show that the number of corrupted input neurons is bounded by a constant times the number of unsatisfied constraints. Thus, driving the number of unsatisfied constraints down to $0$ (which can always be done, as per the previous paragraph) will eventually correct all corrupted neurons (as long as the initial number of unsatisfied constraints is low enough to preclude convergence to the wrong energy minimum).

\section*{Acknowledgments}
We are grateful to David Schwab and Ngoc Tran for many helpful discussions on early parts of this work, and to Yoram Burak and Christopher Hillar for comments on the manuscript. IRF is an ONR Young Investigator (ONR-YIP 26-1302-8750), an HHMI Faculty Scholar, and acknowledges funding from the Simons Foundation. 

\bibliography{chaudhuri_master}

\end{document}


\def\sgn{\mathop{\rm sgn}}

\begin{center}
\textbf{\Huge Supplementary Information}
\end{center}

In order to make this section self-contained, we repeat some portions of the Methods here.

\section{Hopfield networks and Boltzmann machines}
We consider networks of $N$ binary nodes (the neurons). At a given time, $t$, each neuron has state $\vect{x}_i^t$=$0$ or $1$, corresponding to the neuron being inactive or active respectively. The network is defined by an $N$-dimensional vector of biases, $\vect{b}$, and an $N\times N$ symmetric weight matrix $W$. Here $\vect{b}_i$ is the bias (or background input) for the $i$th neuron, and $W_{ij}$ is the interaction strength between neurons $i$ and $j$ (set to 0 when $i=j$).

Neurons update their states asychronously according to the following rule:
\begin{equation}
\vect{x}_i^{t+1}=
\begin{cases}
1 & \text{if } \sum_j W_{ij}\vect{x}_j^t + b_i>0\\
0 & \text{if} \sum_j W_{ij}\vect{x}_j^t +b_i < 0 \\
\text{Bern}(0.5) & \text{if} \sum_j W_{ij}\vect{x}_j^t + b_i = 0
\end{cases}
\end{equation}
Here Bern(0.5) represents a random variable that takes values $0$ and $1$ with equal probability.

Hopfield networks can also be represented by an energy function, defined as
\begin{equation}
E(\vect{x}|\vect{\theta},W)=-\frac{1}{2}\sum_{i\neq j} W_{ij}\vect{x}_i\vect{x}_j - \sum_i \vect{b}_i \vect{x}_i
\label{eq:hopf_energy}
\end{equation}
The dynamical rule is then to change the state of a neuron if doing so decreases the energy (and to change the state with $50\%$ probability if doing so leaves the energy unchanged).

We also consider Boltzmann machines, which are similar to Hopfield networks but have probabilistic update rules.

\begin{align}
\vect{x}_i^{t+1}&=\text{Bern}(p) \nonumber \\
\text{where } p&=\frac{1}{1+e^{-\beta(\sum_j W_{ij}\vect{x}_j^t - b_i)}}
\end{align}
The probability of a state $\vect{x}$ in a Boltzmann machine is $p(\vect{x}) \propto e^{-\beta E(\vect{x})}$, where $E(\vect{x})$ is defined as in Eq. \ref{eq:hopf_energy} and $\beta$ is a scaling constant (often called inverse temperature). Note that the Hopfield network is the $\beta \rightarrow \infty$ limit of a Boltzmann machine.

\section{Error correcting codes}
Given a string of variables (message) to be transmitted, an error correcting code adds redundancy to allow the message to be recovered despite added noise. A parity check code over some set of variables $x_i$, where each $x_i \in \{0,1\}$ is defined by a set of constraints, $\sum x_i=0$, where the sums are taken modulo 2. A classical example of this is the (7,4) Hamming code \citep{Hamming50}, which is defined by considering 4-bit messages and adding 3 parity-check bits to the message, defined as
\begin{align}
x_5&=x_1+x_2+x_3 \nonumber \\
x_6&=x_2+x_3+x_4 \nonumber \\
x_7&=x_1+x_3+x_4
\end{align}

For example, instead of transmitting the message $0101$,  three additional check bits are added on and the message transmitted is $0101101$. Thus there are $2^4$ possible correct messages (see Fig. 2); these possible message are called codewords. If a message is received that does not correspond to a codeword, it is mapped to the closest codeword, thus correcting errors. The parity check bits are chosen so that any two codewords are separated by the state of at least three bits. Thus, if a transmitted message is received where a single bit is flipped, the Hamming code can recover the original message. 

The \emph{distance} of a code is the separation between codewords, and is twice the number of errors that can be corrected. The \emph{rate} of a code is the number of information bits transmitted per message bit. The (7, 4) Hamming code has a distance of $3$ and a rate of $4/7$.

Considering longer blocks of bits allows the construction of codes with better performance, meaning that they either have a larger distance between codewords (i.e., correct more errors) or send information at a higher rate (i.e., more efficiently), or some combination of the above. 

Rather than seeing a codeword as the combination of a desired message and a set of added check bits, the parity check bit equation can be reframed as a set of 3 constraints on 7 variables:
\begin{align}
x_1+x_2+x_3+x_5&=0 \nonumber \\
x_2+x_3+x_4+x_6&=0 \nonumber \\
x_1+x_3+x_4+x_7&=0
\end{align}
The codewords are the states that satisfy these equations. Thus the Hamming codewords occupy a 4 dimensional subspace of a 7 dimensional space.

The constraint structure of a code can be represented as a bipartite graph\footnote{A bipartite graph is a network with two sets of nodes. Nodes in each set connect only with nodes from the other set.}, with one set of variable nodes and another set of constraint nodes. This is shown for the (7,4) Hamming code in Fig. 2, with $7$ variables and $3$ constraint nodes (each corresponding to an equation). Analyzing and constructing codes from a graph-theoretic perspective has been a very fruitful area of research for the last three decades \citep{Tanner81}.

\section{Mapping between general linear codes and higher-order Hopfield networks}
A parity check code can be simply mapped onto Hopfield networks whose nodes, $s_i$ take states in \{-1, +1\} by mapping binary state 0 to +1 and binary state 1 to -1.\footnote{Note that a mapping also exists to Hopfield networks where nodes take states in \{0,1\}, but the energy function is slightly more complex.} The parity check constraints $\sum_{i \in C_i} x_i=0$ can be rexpressed as products $\prod_{i \in C_i} s_i=1$ and then used to define an energy function
\begin{equation}
E(\vect{s})=-\sum_{C_i} \prod_{i \in C_i} s_i
\end{equation}
This energy function takes its minimum energy values if and only if all the constraints are satisfied and the energy increases as the number of the violated constraints increases:
\[
E(\vect{s})=E_{min}+2N_{VC}(\vect{s}),
\]
where $N_{VC}(\vect{s})$ is the number of violated constraints in state $\vect{s}$. Thus the minimum energy states of this network are the codewords of the corresponding parity-check code.

Continuing our Hamming code example, a Hopfield network with the same minimum energy states as the codewords of the (7,4) Hamming code has energy function:
\begin{equation}
E=-s_1s_2s_3s_5-s_2s_3s_4s_6-s_1s_3s_4s_7.
\end{equation}
Note that this involves higher-order edges, meaning edges that connect more than 2 nodes. However, by adding hidden nodes (as we do for our exponential capacity Hopfield network construction later), an equivalent network can be constructed with pairwise interactions.

While the constructed network has the right energy minima (codewords), the dynamics do not do optimal decoding (i.e. error-correction). Optimal decoding would map each corrupted codeword to the most likely original codeword. For IID noise at each variable, this corresponds to the nearest codeword (in Hamming distance, meaning the codeword reached by the fewest variable flips from the current state). By contrast, the Hopfield network energy-based dynamics flips a node if doing so reduces the number of violated constraints (which is proportional to the energy). While the nearest codeword has lower energy than the current state, so do all the other codewords, and energy-based decoding will not necessarily guide the state to the right codeword. Moreover, it may get stuck in local energy minima. Thus while it is easy to write down an energy function that returns the codewords as minimum energy states, most error-correcting codes cannot be decoded by a local dynamical rule. We elaborate on this in the next section.

\section{Energy-based decoding for Hamming and other codes}
For energy-based decoding, we require that the number of violated constraints can serve as a local error signal, meaning that we can decide to flip a node based purely on whether doing so reduces the number of violated constraints. In this section we first prove that such decoding generically fails for Hamming codes, and then provide a heuristic argument for why such decoding requires codes where each variable participates in only a small number of constraints (i.e., the bipartite graph representation is sparse), and where small sets of variables do not share many constraints in common.

\subsection*{Hamming code decoding}
For each value $k\geq 2$, there exists a Hamming code of length $N=2^k-1$ with $k$ constraints. The code conveys $2^k-k-1$ bits of information and can correct one error \citep{Hamming50, Mackay04}. To construct the constraints, first express each variable / message bit in binary (ranging from $1$ to $2^k-1$). Then the $j$th constraint is the sum of all variables that have the $j$-th bit set in their binary expansion. For example, the first constraint sums up bits 1, 3, 5, 7 and so on, and the second constraints sums up bits 2, 3, 6, 7, 10, 11 and so on\footnote{When applied to $k=3$ this yields the $(7,4)$ Hamming code from Figure 2 up to a relabeling of variables}. This construction implies that any given variable has a fixed probability $p$ of participating in each constraint, and $p \approx 1/2$ ($p$ is not exactly half because there is no variable that participates in 0 constraints, but the difference shrinks as N gets larger).

Now consider starting at a codeword and flipping the state of a randomly-chosen variable, which we call $x_i$ (note that all states in a Hamming code are either a codeword or at a Hamming distance of 1 from a codeword). This makes some set $R$ of constraints unsatisfied, where $|R| \sim \text{Bern}(N,p)$ and is $N/2$ on average. To perform energy-based decoding, we consider the effect of flipping a second node, $x_j$, on the number of unsatisfied constraints. We already know that flipping $x_i$ back to its original state will make all constraints satisfied. Flipping any of the remaining nodes will lead the system away from the nearest codeword and to avoid this we would like them to have higher energy. 

If $x_j$ is connected to a set $S$ of constraints, then flipping $x_j$ will change the state of constraints in $S\cap R$ from unsatisfied to satisfied and change those in $S\setminus R$ from satisfied to unsatisfied. Thus the energy of the state with $x_j$ flipped is determined by $|S\setminus R| - |S\cap R|$. Since these sets are chosen independently and with $p=1/2$, on average $|S\setminus R| = |S\cap R|$. At least $50\%$ of possible flips have $|S\cap R|\geq |S\setminus R|$ and these lead to states that have equal or lower energy. Thus about $N/2$ possible directions lead to states that have equal or lower energy, and only $1$ of these leads to the desired codeword. 

Note that here the problem is the overlap between the constraints connected to the variables $x_i$ and $x_j$, suggesting we would like this to be small. Also note that, in this case, gradient descent (i.e., picking the neighboring state with \textit{lowest} energy, not just any state with lower energy) would lead in the right direction, but this is a consequence of our initial state being next to a codeword / global minimum and is not generic. 

\subsection*{Local energy-based decoding of other codes}
We next heuristically argue that for good local energy-based decoding, a code must be sparse (i.e., each variable participates in a sparse subset of constraints) and that variables should not share too many constraints in common.

First consider a code of length $N$ that can correct $O(N)$ errors\footnote{Note that the Hamming code can correct only 1 error.}, and where each variable participates in a fraction $pN$ of constraints. Start at a codeword and flip a set of $\alpha N$ bits, which we call $E$. In the absence of special structure, $E$ contacts a number of constraints that grows as $N^2$, and thus even for very small $\alpha$ and $p$, all possible constraints will be connected and will receive multiple edges from the nodes in $E$. Thus their states will be approximately random. Now flipping a node that is outside of $E$ will change the state of a set of $pN$ constraints, which we call $T$. Satisfied constraints in $S$ will change to unsatisfied and vice versa. The new state will have lower energy if $T$ contains more unsatisfied than satisfied constraints, which will happen with about $50\%$ probability. Thus there are many variables outside of $E$ that lead to lower energy states, and energy-based decoding will in general not recover the nearest codeword.

Next, consider codes of length $N$ where each variable participates in a small number of constraints that does not grow with $N$. As before, consider a set of error nodes $E$, which will be connected to some small set of constraints, $S$. Some subset of constraints in $S$ are unsatisfied. For good energy-based decoding, we would like flipping variables outside of $E$ to increase the number of unsatisfied constraints, and flipping variables in $E$ to decrease the number of unsatisfied constraints. In both cases, this is determined by the overlap of connected constraints between the variable to flip and the nodes in $E$. First, consider flipping a node $x_i$ outside $E$. $x_i$ is connected to some set of constraints $T$. All constraints in $T\setminus S$ will switch from satisfied to unsatisfied, and some subset of constraints in $T \cap S$ will switch from unsatisfied to satisfied. Thus, we would like $T \cap S$ to be as small as possible (and $T\setminus S$ to be large). Similarly, consider flipping a node $x_i$ inside $E$, which is connected to some set of constraints $T \subset S$. Constraints in $T$ that are only connected to $x_i$ and not to other members of $E$ will become satisfied, while a subset of constraints in $T$ that receive multiple edges from $E$ will become unsatisfied. As before, we wish this subset to be small and the number of constraints in $T$ that are only connected to $x_i$ to be large. Thus we wish the overlap of constraints between $x_i$ and $E \setminus x_i$ to be small. This argument suggests that codes decodable by a local energy-based rule should be sparse and that small sets of variables should not share many constraints in common.

\section{Expander graphs}
Expansion is a property of a graph (i.e., network) where small sets of nodes have a large number of neighbors (i.e., connected nodes). In the context of coding theory, sparse expander graphs allow variables to not share many constraints in common, allowing for local energy-based decoding.

There are various ways to formalize the notion of expansion. We consider bipartite graphs, meaning graphs that are divided into two sets of nodes, with connections between these sets but no connections within a set. Consider an undirected bipartite graph with $N$ nodes in an input layer and $N_C \sim N$ nodes in a hidden layer, which we call a {\em constraint} layer. Assume that input nodes have connections drawn from some degree distribution with degree $z$ such that $z^{min} \leq z \leq z^{max}$. Similarly, constraint nodes are drawn from a distribution with $z_C<z_C^{max}$. Such a graph is a $(\gamma, (1-\epsilon))$ expander if every set of nodes $S$ in the input layer with $|S|\leq\gamma N$ has at least $(1-\epsilon)|\delta(S)|$ neighbors, where $\delta(S)$ is the set of edges connected to nodes in S, and $|\delta(S)|$ is the number of edges in this set. Thus small subsets of variables (``small'' is determined by $\gamma$) participate in proportionately large sets of constraints (``large'' is determined by $1-\epsilon$). Note that if the edges emerging from $S$ target disjoint nodes, then $\epsilon=0$. Thus, $\epsilon \rightarrow 0$ corresponds to increasing expansion.

Expander graphs can be constructed in various ways, but sparse random bipartite graphs are generically expander graphs \citep{Sipser96, Luby01}. We use the following lemma from Luby et al. (2001).

\begin{lem}
Let $B$ be a bipartite graph, with nodes divided into $N$ left nodes and $N_C$ right nodes. Suppose that a degree is assigned to each node so that all left nodes have degree at least five, and all right nodes have degree at most $C$ for some constant $C$. Suppose that a random permutation is chosen and used to match each edge out of a left node with an edge into a right node. Then, with probability $1-O(1/N)$, for some fixed $\gamma>0$ and $\epsilon<1/4$, $B$ is a $(\gamma, (1-\epsilon))$ expander.
\label{lem: random_expander}
\end{lem}
In our simulations we generate all graphs randomly by picking edges to connect pairs of variable and constraint nodes subject to the constraints on the degree distributions.

\begin{figure}
    \begin{center}
        \includegraphics[width=0.5\linewidth]{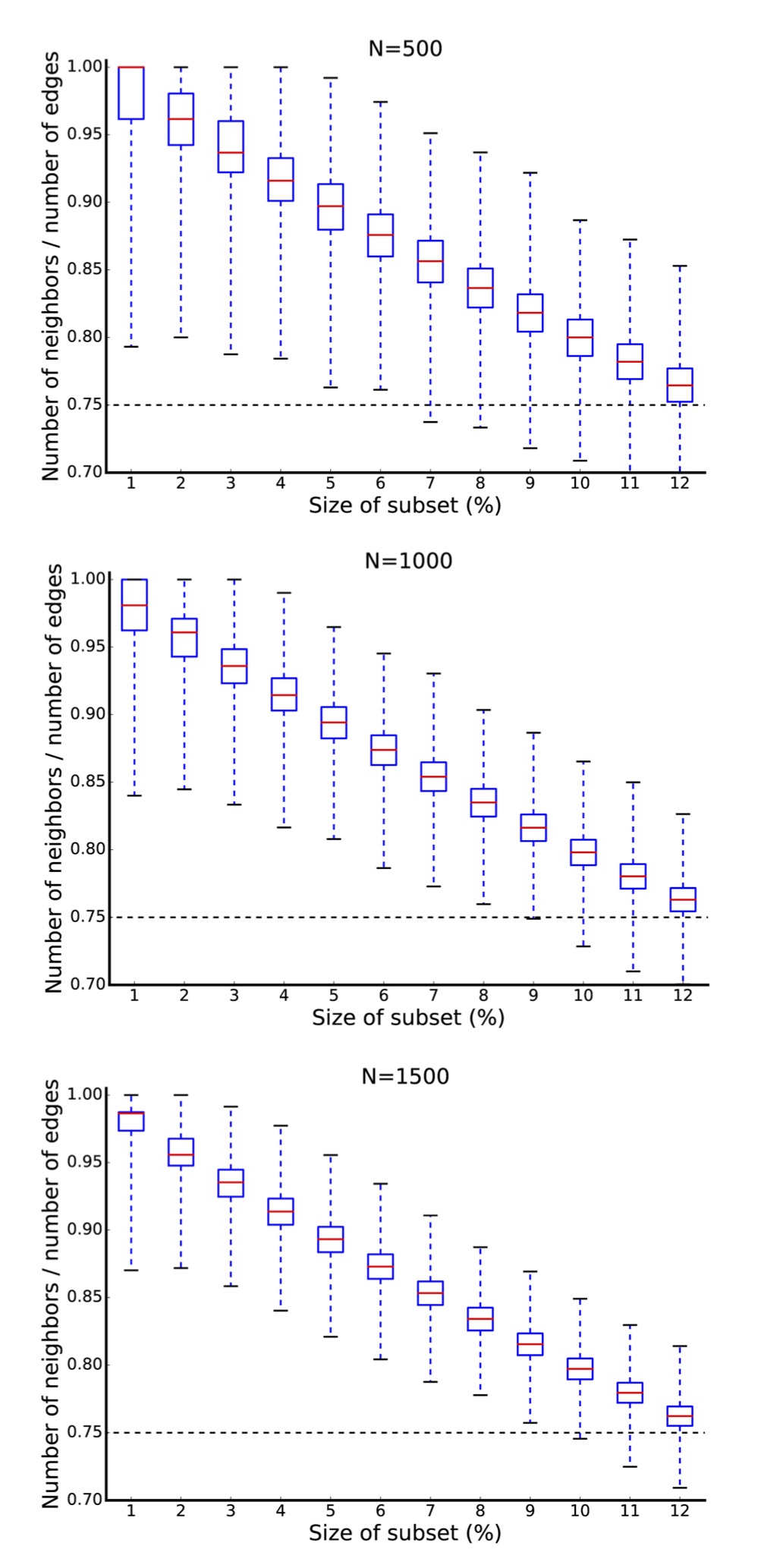}
    \end{center}
    \caption{{\bf Estimates of network expansion.} 
Each plot shows the ratio of the number of neighbors to the number of edges for sample subsets of nodes drawn from the networks used to generate Figure 3. This ratio corresponds to $(1-\epsilon)$ in Eq. 4 in the main text. Dashed line shows $(1-\epsilon)=0.75$, which is the theoretical lower bound on expansion required for good error correction.}
\label{fig:expansion}
\end{figure}

Also note that the Hamming codes we considered previously have expansion that goes to $0$\footnote{For example, in a Hamming code with $N$ variables and $k$ constraints, consider the set $\hat{S}$ of variables that participate in $k-1$ constraints. There are $k$ such variables, and each sends out $k-1$ edges. The set as a whole sends out $|\delta(\hat{S})| = (k-1)k$ edges and is connected to every constraint node so has $|N(\hat{S})| = k$ neighbors. Consequently, $|N(\hat{S})|/|\delta(\hat{S})| \to 0$ as $k \to \infty$. For expansion, we require that for some $\gamma>0$, any set $S$ of size $|S|<\gamma N$ has at least $(1-\epsilon)|\delta(S)|$ neighbors. The set $\hat{S}$ has size $O(\log(N))$ and thus, no matter how small $\gamma$ is, for large enough $N$ we can choose a set of size $<\gamma N$ that contains $\hat{S}$. This set has $|\delta(S)| \geq |\delta(\hat{S})|$ and $|N(S)| = |N(\hat{S})|$ (since $\hat{S}$ is connected to every constraint). Consequently, for all $\gamma>0$, $|N(S)|/|\delta(S)| \to 0$ as $k \to \infty$.}.
  
\section{Irregular expander codes with general constraints}
\label{sec:bit_flip_decoding}

The following analysis is based on Sipser \& Spielman \citep{Sipser96} and Luby et al. \citep{Luby01}, with slight generalization to consider the case of general rather than parity constraints. We consider an undirected bipartite graph with $N$ nodes in the input layer and $N_C \sim N$ nodes in the hidden layer, which we call a {\em constraint} layer. Assume that input nodes have connections drawn from some degree distribution with degree $z$ such that $z^{min} \leq z \leq z^{max}$. Similarly, constraint nodes are drawn from a distribution with $z_C<z_C^{max}$. We assume that such a graph is an expander for some $\gamma>0$ and with $\epsilon<1/4$. Note that the lemma in the previous section guarantees that such graphs can be constructed randomly.

To prove that the network performs good decoding / pattern completion, we consider a network where the input nodes are in state $Q$, where $Q$ differs from some satisfied state $Q_0$ on a set of nodes $E$. We will show that if $Q$ lies within some distance $d$ of $Q_0$ (where distance=$|E|$ is the number of nodes on which the two states differ), then the network dynamics converges to $Q_0$.

Consider the set of error nodes, $E$, and assume that $|E|\leq\gamma N$. Define $U(E)$ to be the unsatisfied constraints connected to $E$ and $S(E)$ to be the satisfied constraints connected to $E$. The neighbors of $E$ are $N(E)=U(E) \cup S(E)$ (and the number of neighbors is $|N(E)|$). Define a {\em unique neighbor} of $E$ to be a constraint node that is only connected to one node in $E$, and $\text{Unique}(E)$ to be the set of such neighbors. We start by lower bounding $|\text{Unique}(E)|$ using a counting argument.

Since $|E|\leq\gamma N$, the expansion property holds, and $E$ has at least $(1-\epsilon)|\delta(E)|$ neighbors. Thus, at least $(1-\epsilon)|\delta(E)|$ of the edges in $\delta(E)$ go to different constraint nodes. There are $\epsilon |\delta(E)|$ remaining edges, meaning that at most $\epsilon |\delta(E)|$ constraint nodes can receive more than one edge and the remainder receive exactly one edge from $E$ and are unique neighbors. Consequently
\begin{equation}
|\text{Unique}(E)|\geq (1-\epsilon)|\delta(E)|-\epsilon |\delta(E)|=(1-2\epsilon) |\delta(E)|.
\label{eq:unique_nbrs}
\end{equation}

The number of unique neighbors of $E$ is determined purely by graph connectivity and not by the particular constraints imposed.

Next, we translate the $|\text{Unique}(E)|$ into a bound on $U(S)$, the set of unsatisfied constraints. In the simplest case, the acceptable states for each constraint differ on at least 2 variable nodes (we weaken this assumption in the following section). Any constraint $C \in \text{Unique}(E)$ is connected to only one corrupted variable node and is thus violated. Consequently,
\begin{equation}
|U(E)|\geq(1-2\epsilon) |\delta(E)|.
\label{eq:unsat_constraints_min_sep}
\end{equation}
Note that provided $\epsilon>1/2$, this guarantees that any states which satisfy all constraints must differ on $>\gamma N$ nodes.

The randomized construction allows us to construct networks with $\epsilon>1/4$, guaranteeing that $|U(E)|>|\delta(E)|/2$. Thus at least half the edges leaving $E$ target unsatisfied constraints, meaning that at least one node in $E$ is adjacent to more unsatisfied than satisfied constraints. Hence there is always a node whose state is energetically unfavorable and that will eventually switch.

This statement is stronger than the claim that states with satisfied constraints are separated by $\gamma N$. While states with all satisfied constraints are minima, it might have been the case that there also exist local minima with unsatisfied constraints within this $\gamma N$ radius. For example, if all the error nodes were connected to more satisfied than unsatisfied constraints, then flipping any single node would increase the energy function. The corresponding input state would be a local minimum and would be an acceptable steady-state for the neural network (a non-local decoding algorithm could circumvent this by searching over a wider set of neighbors). Note that local minima might still exist, but not within a radius of $\gamma N$ of a state with all satisfied constraints.

Thus far we have shown that minima of the energy function must differ on the states of $>\gamma N$ variables but not that these minima have appropriate basins of attraction. To establish this, consider a network state on decoding step $t$, $Q(t)$, which differs from some state with all constraints satisfied, $Q_0$, on a set of nodes $E(t)$. Let $U(t)$ be the set of unsatisfied constraints at time $t$, and note that the network dynamics always decreases the number of unsatisfied constraints, so $U(t+1)<U(t)<\cdots<U(0)$. 

We require that $|E(0)|<\frac{z^{min}}{z^{max}}\frac{\gamma N}{2}$ (recall that $z^{min}$ and $z^{max}$ bound the degree of nodes in the input layer). Each variable in $E$ sends out a maximum of $z^{max}$ edges, and a constraint node can only be unsatisfied if it receives one of these edges. Thus 
\begin{equation}
|U(t)|\leq|U(0)|\leq|\delta(E(0))|\leq z^{max}|E(0)|<z^{min}\frac{\gamma N}{2}
\label{eq:unsat_num_bound}
\end{equation}

On the other hand, energy minima are separated by a distance of at least $\gamma N$. Thus if the network starts in a state with distance $|E|<\gamma N$ from $Q_0$ and ends up at another energy minimum, it must pass through a state with $|E|=\gamma N$. By Eq. \ref{eq:unsat_constraints_min_sep}, this intermediate state has at least $(1-2\epsilon) |\delta(E)|\geq \frac{|\delta(E)|}{2}\geq z^{min}\frac{\gamma N}{2}$ violated constraints, which violates Eq. \ref{eq:unsat_num_bound}.

In summary, while the network dynamics may (transiently) increase the size of $E$, it does not increase the number of violated constraints: if we start in a state with fewer than $\frac{z^{min}}{z^{max}}\frac{\gamma N}{2}$ errors we will always remain in a state with fewer than $\gamma N$ errors. In this case Eq. \ref{eq:unsat_constraints_min_sep} guarantees that there is always a node which will change its state, and doing so reduces the number of violated constraints by at least one. This means that the network will converge in a time bounded by the product of the number of violated constraints and the time it takes each variable node to flip.

\section{Hopfield network expander codes: construction}
We now construct Hopfield networks that implement the dynamics of the error-correcting codes defined in the previous section. Before we consider networks with pairwise connectivity, note that if higher-order edges are allowed (as in S3), then Hopfield networks with sparse random connectivity are isomorphic to expander codes (each parity constraint corresponds to a higher-order edge), and thus generically have exponential capacity and large basins of attraction.

We now consider networks with pairwise connectivity. As in the previous section, the networks we consider are bipartite, containing $N$ input nodes, which determine the states or memories that will be stored and corrected, and $N_C$ constraint nodes, which determine the allowed states of variables they are connected to (see Figure 4). However, now these constraint nodes are themselves small networks of neurons (see below). As before, the $i$th input neuron connects to $z^{(i)}$ constraint nodes, and the $j$th constraint node connects to $z_C^{(j)}$ inputs. Consequently $\sum_i z^{(i)} = \sum_j z_C^{(j)}$ (i.e., the number of edges leaving the input nodes equals the number of edges entering the constraint nodes). $z$ and $z_C$ are small and chosen from distributions that do not scale with $N$; consequently the networks are sparse. For the simulations in Figure 3, we set $N_C = 0.95 N$, and choose $Z = 4 + Z_{add}$, where $P(Z_{add} = k) = 0.85 \times 0.15^{k-1}$ (note this is a geometric distribution with $p=0.85$). We then randomly assign outgoing edges from variable nodes to constraint nodes, subject to $2\leq Z_C\leq 6$. In Figure 4, we choose deterministic values of $z=5$ and $z_C=12$ (and consequently $N_C = 5N/12$). By Lemma \ref{lem: random_expander}, these networks are $(\gamma, (1-\epsilon))$ expanders with $\epsilon<1/4$.

There are $z_C$ variables connected to a given constraint node and these could take any of $2^{z_C}$ possible states. The constraint nodes restrict this range, so that a subset of these states have low energy (and are thus preferred by the network). Each constraint node is actually a network of neurons with Hopfield dynamics and, while there are multiple possible ways to construct constraint nodes, in the constructions we show each neuron in a constraint node prefers one possible configuration of the variable nodes. For Figure 3 we choose these configurations to be parity states of the connected input nodes (this simplifies the numerical simulations, but is not necessary), and in Figure 5 we choose these configurations randomly, but subject to the constraint that each preferred configuration differs from the others in the state of at least two neurons. 

If a neuron in a constraint node prefers a particular state of its input nodes, then it has connection weights of +1 with the nodes it prefers to be active and weights of -1 with the nodes it prefers to be inactive (see Figure 3 for an illustration). Thus, a constraint neuron that prefers all of its inputs to be active will receive input of $z_C$ when its input is in its preferred state, and a constraint neuron that prefers only $k$ of these nodes to be active will receive input of $k$. To ensure that all constraint neurons receive the same amount of input in their preferred state, we also add biases of $z_C-k$ to each constraint neuron, where $k$ is the number of non-zero variables in its preferred state. Finally, to ensure that multiple constraint neurons are not simultaneously active, we add inhibitory connections to strength $z_C-1$ between all the neurons in a constraint network. As a consequence of this strong inhibition, each constraint network has competitive dynamics: in the lowest energy state the variable nodes are in a preferred configuration, the neuron in the constraint network corresponding to this configuration is active, and all other neurons are suppressed. Note that this recurrent inhition is not necessary and can be replaced by non-specific strong background inhibition to all constraint neurons, though this slows down the network convergence.

The network is in a stable or minimum energy state when all of the constraints are satisfied. If each constraint is satisfied by a fraction $2^{-r_j}$ of possible states, where $r_j \geq 1$ but is not necessarily integer, then the average number of minimum energy states for the network is
\[
N_{\text{states}}=2^{-\langle r\rangle N_C}2^{N}=2^{N-\langle r\rangle N_C}=2^{\left(1-\langle r\rangle \frac{\langle z \rangle }{\langle z_C \rangle}\right)N}.
\]
Here the angle brackets represent averages, and the last equality follows because $N\langle z \rangle=N_C \langle z_C \rangle$. For notational convenience define $\hat{z} = \frac{\langle z \rangle }{\langle z_C \rangle}$. Thus, as long as $\langle r\rangle \hat{z}<1$, the expected number of stable states grows exponentially with $N$.

However, $N$ is the number of input nodes and not the total network size. Each constraint network has at most $M=2^{z^{max}_C-1}$ neurons, and thus the total number of nodes in the network is at most $N_{net} \leq N+MN_C=(1+M\hat{z})N$. Since $z$ and $z_C$ are drawn from fixed distributions, this prefactor does not grow with network size. The number of minimum energy states is
\begin{equation}
N_{\text{states}} \geq 2^{\alpha N_{net}} \quad \text{where} \quad \alpha=(1-\mean{r}\hat{z})/(1+M\hat{z})
\end{equation}

Thus the number of minimum energy states grows exponentially in the total size of the network.

\section{Hopfield network expander codes: error-correcting dynamics}
We show that the Hopfield network dynamics carries out the decoding algorithm described in Section \ref{sec:bit_flip_decoding}.

For simplicity, we first consider the case where the input neurons are clamped to some fixed state. Some constraint nodes are satisfied, meaning that the input neurons they are connected to take one of the allowable states of the constraint. The constraint neurons in a satisfied constraint node settle down to a state where the neuron that prefers the particular input is active and the other are inactive. Other constraint nodes are unsatisfied, meaning that their input neurons take a non-preferred state. In this case, no neuron in the constraint node prefers the current input state, but some of them prefer a neighboring input configuration and are thus weakly driven. Because of the strong recurrent inhibition, the node settles into a state where either one or two of these weakly-driven neurons are active, and the node drifts between all such sparse combinations of these weakly-prefered neuron. Note that there may only be one such weakly-preferred neuron, in which case this neuron will be active.

Now consider the node as a whole. Consider an input neuron, X, that is connected to S satisfied and U unsatisfied constraints. If $S>U$ then X will not switch state because doing so increases energy (see below for networks that tolerate some small probability to move to higher energy states). However, if $U>S$ then it is sometimes energetically favorable to switch. Upon switching, the constraint nodes in S will become unsatisfied. On the other hand, the neurons in the unsatisfied constraint nodes are wandering between states that prefer various neighbors of their current state, and switching X will satisfy some of them but will further increase energy at the others. Consequently, input neurons switch slowly but, since the size of the constraint nodes does not grow with network size, this time does not grow with network size. By contrast, neurons in constraint nodes switch their states quickly and rapidly settle to equilibrium with the input nodes. Thus the network implements the bit-flip dynamics of Section \ref{sec:bit_flip_decoding} and, consequently, can correct a number of errors that scales with network size.

\section{Weakening constraints}
In the previous section we considered constraints with a guaranteed minimum distance of $2$ between their satisfied states. We now extend those results to consider weaker constraints that accept some fraction of adjacent states. For example, such a constraint might be satisfied when its connected variables take configurations $(0,0,0)$, $(0,0,1)$, and $(1,1,1)$. We define $p$ to be the probability that a neighbor of a satisfied state is unsatisfied. We show that for slightly higher expansion, the results above will hold on average, and we bound the deviation from this average for large $N$.

As before, consider a state $Q$ differing from a satisfied state $Q_0$ on nodes $E$. In the previous section we had $\text{Unique}(E)\subset U(E)$ (and, implicitly, $p=1$). However, in the weaker setting, a constraint in $\text{Unique}(E)$ is unsatisfied with probability $p$. For notational convenience define the random variable $X=|U(E)|$ (i.e., the number of unsatisfied constraints), and note that $X \geq \text{Binomial}(K,p)$, where $K=|\text{Unique}(E)|\geq(1-2\epsilon) |\delta(E)|$. As before, we wish $X>|\delta(E)|/2$.

For this to hold we will require higher expansion. Choose $\epsilon$ so that $\frac{1}{2(1-2\epsilon)}<p$ (if $p=1$ then $\epsilon<1/4$ as before). Now define $\alpha=\frac{1}{2p(1-2\epsilon)}<1$ and note that $\mathbb{E}[X]\geq pK> |\delta(E)|/(2\alpha)>|\delta(E)|/2$. Thus, on average, the state $Q$ will have a node that it is energetically favorable to flip.

We now bound the probability of error, meaning the probability that the number of unsatisfied constraints is less than half the number of edges leaving $E$.

\begin{align}
P\left(X < \frac{|\delta(E)|}{2}\right)&\leq P\big(X< \alpha \mathbb{E}[X]\big) \nonumber \\
&\leq \exp \left(-\frac{(1-\alpha)^2}{2}\mathbb{E}[X]\right) \nonumber \\
&< \exp \left(-\frac{(1-\alpha)^2}{4\alpha}|\delta(E)|\right).
\label{eq:weak_constraints_error_prob}
\end{align}
Here we use the Chernoff bound for the second inequality.

$|\delta(E)|$ is approximately proportional to the number of error variables, $|E|$ (for a fixed degree network this is exact and the proportionality constant is just the degree). Thus, the probability that there exists another minimum within a distance $d$ of $Q_0$ falls off exponentially in $d$. 

These results allow the presence of other local minima, which we can divide into two categories. First, there may be a small number of local minima very close to $Q_0$, at a distance that does not scale with network size (and thus a distance that vanishes in relative terms). The effect of these minima is to slightly expand the desired energy minima to possibly include a set of nearby states rather than a single state, but the size of this set does not grow with network size or number of minima. Second, while the probability that there exists another minimum within a distance $d$ of $Q_0$ falls off exponentially in $d$, the number of states at distance $d$ grows exponentially in $d$, and thus there will be $O(1)$ local minima at distance $d$. However, since these minima are produced by rare events, the basins of attraction are likely to be small and most trajectories should not see these minima.

Next, we note that the probability of these local minima decreases exponentially in the variable degree. To see this, consider Eq. \ref{eq:weak_constraints_error_prob} for a fixed degree network, where each variable particiaptes in $z$ constraints. Then $|\delta(E)|=z|E|$,
\begin{align}
P(\text{local minimum}) &< \exp \left(-\frac{(1-\alpha)^2}{4\alpha}|\delta(E)|\right) \nonumber \\
&= \exp \left(-\frac{(1-\alpha)^2}{4\alpha}z|E|\right).
\end{align}

Thus it is always possible to achieve an error below any given fixed probability by choosing $z$ appropriately and this value of $z$ does not need to grow with network size. Moreover, the error probability can be asymptotically driven to $0$ by allowing $z$ to grow with $N$ at any rate.

\section{Noisy updates / finite temperature}
\begin{figure}[h!]
    \begin{center}
        \includegraphics[width=0.8\linewidth]{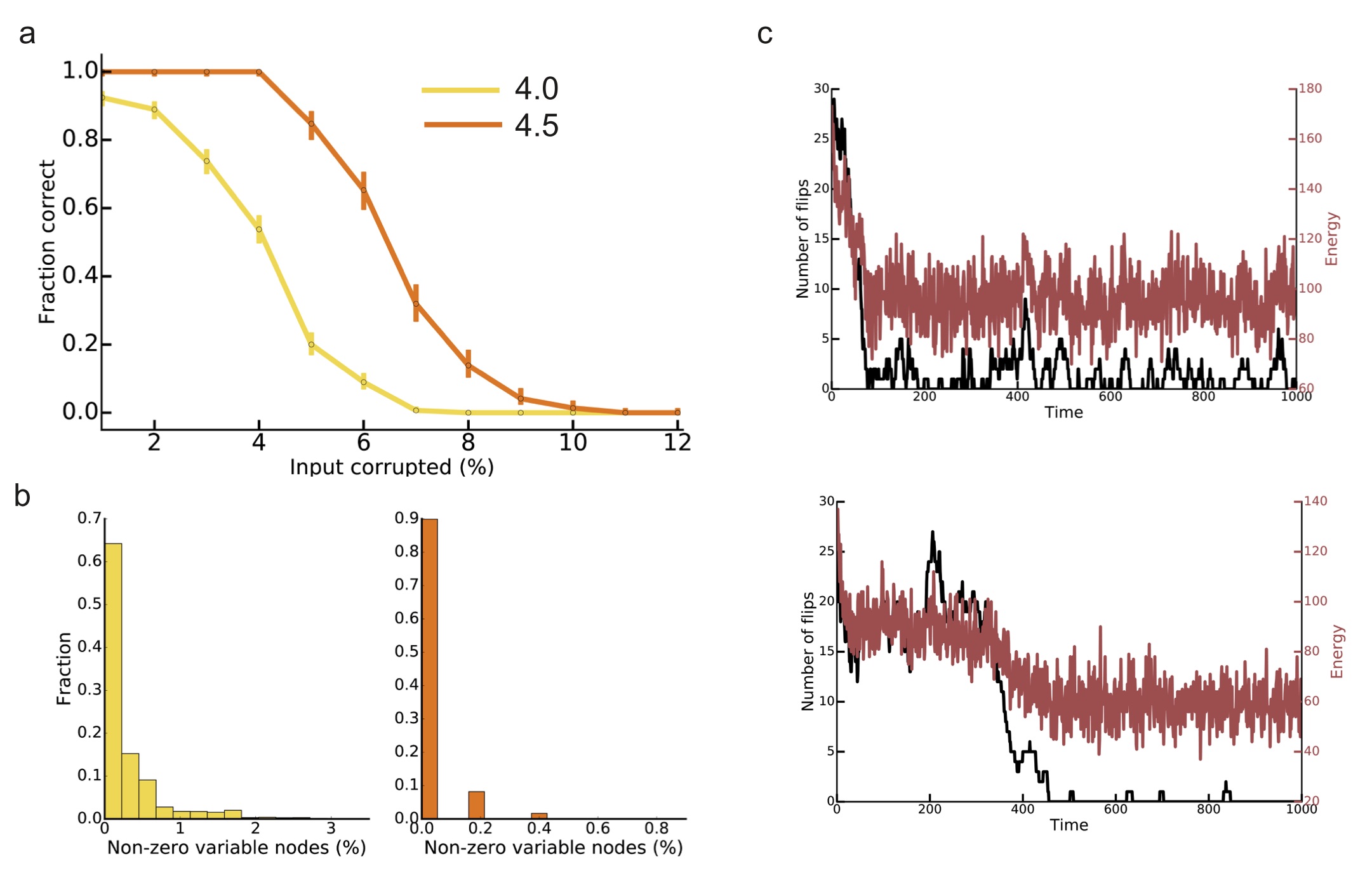}
    \end{center}
    \caption{{\bf Network dynamics at finite temperature.} 
(a) Fraction of times the network infers the correct state as a function of percent input corrupted, for two different inverse temperatures. The network is considered to have reached the correct state if the final state is within two standard deviations of the mean of the equilibrium distribution shown in \textbf{b}.
(b) Distribution of number of nonzero variable nodes for networks started at the all-zero energy minimum and allowed to evolve. Note that hidden nodes are subject to noisy updates as well (state not shown). Colors correspond to \textbf{a}.
(c) Sample trajectories for two values of inverse temperature. Top panel shows inverse temperature = 4.0 and bottom panel shows inverse temperature = 4.5. Network size of 500 variables in all simulations.}
\label{fig:finite_temp}
\end{figure}
Previously, we established the existence of an energy gradient allowing error correction with large basins of attraction. We now consider the case when input nodes update their state probabilistically rather than always descending the energy gradient. We show that the network state evolves towards the energy minimum on average and, as before, we bound the deviation from this average for large $N$. Note that in this case we do not expect perfect decoding. If we start the network at an energy minimum and there is some small probability $p$ of a node flipping to a higher energy state, then on average $p N$ nodes will flip, and thus the distribution of network states will be localized around but not exactly at the minimum.

Consider a node deciding which state to take, with $Q$ and $Q^\sharp$ the possible network states (differing only on the value of that node). Assume that $P(Q)=1-P(Q^\sharp)=f(\Delta E(Q,Q^\sharp))$, where $\Delta E$ is the energy difference between the two states and $f$ is some function. For a Hopfield network this function always picks the lower energy state, while for a Boltzmann machine the ratio of probabilities for the two states is exponential in the energy difference.

We consider the basin of attraction around a minimum energy state $Q_0$ and consider some general state $Q$ in this basin of attraction. The energy of state $Q$, $E(Q)$ is proportional to $N_{VC}$, the number of violated constraints, with some constant $k_1$ (the constant is irrelevant for the Hopfield network formulation but not when the switching probability depends on the energy difference, such as with a Boltzmann machine). If $F$ is the set of error locations, then the previous analyses show that the number of violated constraints, $N_{VC}\geq(1-2\epsilon)|\delta(F)|$. Moreover, no constraint is violated unless it receives at least one edge from a variable in $F$. Thus $N_{VC}\leq |\delta(F)|$. For simplicity, we consider the case when the degrees of the nodes are constant, so that $|\delta(F)|=z|F|$ (but note that we're considering sets whose size scales with $N$, and thus $|\delta(F)|$ will be increasingly concentrated around $\langle z \rangle |F|$ in the large $N$ limit). Combining,
\begin{equation}
k_1 (1-2\epsilon)z|F|\leq E(Q) \leq k_1 z |F|
\end{equation}

We make the approximation that $E(Q) \propto |F|$. Thus increasing the size of $|F|$ by $1$ changes the energy by some fixed value $\Delta(E)>0$ and decreasing $|F|$ changes energy by $-\Delta(E)$. Consequently, each node in $Q$ takes the same state as it does in $Q_0$ with probability $1-p$ and takes opposite state with probability $p$, where $p$ is small and depends on $\Delta(E)$.

In equilibrium, the average size of $F$ will be $pN$ and, for large $N$, fluctuations around this will be on the order of $\sqrt{N}$. Recall that the basins of attraction have size $\frac{\gamma N}{2}$. Thus in order for decoding to work we require that $p<\frac{\gamma}{2}$.

We analyze the effect of updating a state $Q$ where $|F|=\alpha Q$, meaning that a fraction $\alpha$ of nodes have a state different from the energy minimum. We assume $p<\alpha$, since $pN$ is the best decoding possible and that $\alpha<\frac{\gamma}{2}$, so as to keep the state within the basin of attraction. We show that updates send the network towards state $Q_0$ with high probability.

Consider the network after $M=\phi_{update} N$ nodes have been updated. If nodes update their states in parallel, as typically assumed, then this happens in constant time; if they instead update sequentially then we are effectively considering some constant fraction of the time the network takes to converge. Let the random variable $X_i^{old}$ take value $1$ if the $i$-th node in the update set is in error (i.e. differs from its state in $Q_0$) and $0$ otherwise. Similarly, $X_i^{new}$ is the corresponding random variable after the update. Initially the number of error nodes in this set is $X^{old}=\sum_{i=1}^M X_i^{old} \sim Bin(M, \alpha)$. After the update the number of error nodes in this set is $X^{new}=\sum_{i=1}^M X_i^{new} \sim Bin(M, p)$. Consequently, the set $|F|$ changes in size by $\Delta F=X_{error,new}-X_{error,old}$, and we wish to show $P(\Delta F<0)$ vanishes as $N$ gets large.

Note that $\Delta F=\sum_i Y_i$, where $Y_i=X_i^{new}-X_i^{old}$. Also, $\mathbb{E}[\Delta F]=(p-\alpha)N$. Applying Hoeffding's inequality we find
\begin{align}
P(\Delta F\geq 0)&=P\left((\Delta F-\mathbb{E}[\Delta F])>\frac{(\alpha-p)}{N}\right) \nonumber \\
&\leq \exp \left(-\frac{(\alpha-p)^2\phi N}{2}\right)
\end{align}

Thus this probability decreases exponentially in $N$. Note that the average step size $\Delta F$ gets smaller as $\alpha \to p$.

For the simulations shown in Fig. S2, we scale all the connections described in Section 6 by an inverse temperature $\beta$, and update nodes according to Boltzmann dynamics (Eq. 3).

\section{Self-organization to exponential capacity}
We first describe the learning at a single constraint node. Consider a constraint node that receives $z_C$ inputs from variable nodes, with initially negligible, unstructured connections. We assume that the constraint node has $M$ neurons, where $M\geq 2^{z_C-1}$. Each of these $M$ constraint neurons inhibits the others with recurrent inhibition of strength $\xi$ and receives background inhibition of strength $\eta$. On each learning step, we provide a random input to the $z_C$ input neurons and allow the constraint neurons to reach equilibrium with the inputs. If the input is within a Hamming distance of $1$ from a previously learned state then, provided $|\eta| < (z_C-1)$, the constraint neuron corresponding to that state activates, otherwise all constraint neurons remain inactive. We then provide a random excitatory input of strength $\zeta>|\eta|$ to a randomly selected constraint neuron. If no other constraint neuron is active (i.e., the state has not been learned before), then the neuron that receives this excitatory input activates, and we learn connection strengths of $+1$ with input neurons that are active and $-1$ with input neurons that are inactive, and we add a bias that ensures that the total input drive is $z_C$. If another constraint neuron is active (i.e., the state has been learned before) then, provided that $|\zeta|<|\xi|+|\eta|$, this constraint neuron suppresses the activation of other constraint neurons via the recurrent inhibition and no learning takes place. At the end of learning, we remove the inhibitory bias $\eta$. For the simulations we show, we use $\xi = -(z_C-1)$, $\eta = -(z_C - 1.5)$ and $\zeta = z_C-1$. Note that the learning procedure just described is equivalent to randomly selecting satisfied constraint states from the $2^{z_C}$ possible inputs, subject to the constraint that each new selected state must be a Hamming distance of at least $2$ from previously selected states. 

For the simulations shown in Fig. 5, we start with a randomly-constructed bipartite graph, with degree $5$ at the input nodes and $12$ at the constraint nodes. Thus each input connects to $5$ constraint nodes, and each constraint nodes constrains the state of $12$ variables. We then select acceptable input states randomly for each constraint node, subject to a minimum Hamming distance of $2$ between selected states. 

\section{Retrieving labels for noisy input patterns}
In this section we present a construction that allows the very large number of robust memory states to be used as a neural pattern labeller, in which distributed input patterns are assigned abstract indices corresponding to the memory states. Note that this construction is not specific to our network and can be used for any high-capacity network construction.

Consider a set of $N_{{patt}}$ input patterns, $\{x_i\}$, living in an $M$-dimensional space, which we wish to map to a corresponding set of output patterns, $\{y_i\}$, living in an N-dimensional space. The $y_i$ output patterns are the memory states of a high-capacity memory network, while the $x_i$ are arbitrary. For convenience we will choose values for both sets of patterns from $\{-1, +1\}$ (but note that a simple linear transformation maps these to $\{0,+1\}$). We will assume that $M>>N$, and that $N_{patt}\sim O(M)$. For concreteness, let $N_{patt}=\phi M$, where $\phi<1$. 

We wish to construct an input mapping $U$ that maps each of the $x_i$ input states to the neighborhood of the corresponding $y_i$ states: $Ux_i \approx y_i$ (note that because of the error correcting dynamics in the memory network, the mapping does not need to be exact). Moreover, we wish this mapping to preserve the neighborhood structure, so that slightly perturbed values of $x_i$ are mapped to slightly perturbed values of $y_i$ (again, the error correcting dynamics in the memory network will then recover $y_i$).

In what follows we will consider random input patterns $x_i$. Note that $x_i^Tx_i = M$ and, for $i \neq j$, we can treat $x_i^T x_j$ as a random variable, $X^{prod}_{ij}$, with mean $0$ and variance $M$. 

The memory states $y_i$ are not random. While they have higher-order structure (imposed by the constraints), they are decorrelated to second-order (this is needed for high capacity) and thus, as for the random input patterns, $y_i^Ty_i = N$ and, for $i \neq j$, we can treat $y_i^T y_j$ as a random variable, $Y^{prod}_{ij}$  with mean $0$ and variance $N$.

\subsection*{Outer product construction for $U$}
Define $U_{outer} = \sum_{j=1}^{N_{patt}} y_j x_j^T$ (this can be normalized by $N_{patt}$ if need be). Then $U_{outer}x_i = \sum_{j=1}^{N_{patt}} y_j x_j^Tx_i = My_i + \sum_{j \neq i}y_j X^{prod}_{ij}$ (note that this can be constructed online, given each new pair $x_i$, $y_i$). Now consider the $k$th element of $U_{outer}x_i$ and assume that the $k$th element of $y_i$ (which we call $y_i^k$) is $+1$ (the argument is identical with opposite sign if $y_i^k = -1$). 
\[[U_{outer}x_i]^k = M + \sum_{j \neq i}y^k_j X^{prod}_{ij}.\]

Each term in the sum is a random variable with mean $0$ and variance $M$ (since $y^k_j$ has variance $1$ and is independent of $X^{prod}_{ij}$). Thus $\sum_{j \neq i}y^k_j X^{prod}_{ij}$ has mean $0$ and standard deviation $\sqrt{MN_{patt}}$. If $N_{patt} =\phi M$ then the standard deviation is $\sqrt{\phi}M$. $[U_{outer}x_i]^k$ will have the wrong sign if the contribution from this noise term outweighs the contribution from $y_i^kx_i^Tx_i=M$. Since the standard deviation is the same order as $M$, this will have constant probability at each node, and thus $U_{outer}x_i = y_i + \Delta y_i$, where $\Delta y_i$ has $O(N)$ non-zero terms. Consequently, $U_{outer}x_i$ is mapped to a slightly perturbed value of $y_i$, with the number of errors depending on $\phi$, the number of patterns divided by the size of the input network. 

Note that if we instead choose $N_{patt}$ to scale sub-linearly with $M$ (e.g., $M^\beta$ with $\beta<1$ or $\log(M)$), then the probability of error will decrease with $M$ and the mapping will be asymptotically exact.

We next show that slightly perturbed values of $x_i$ are mapped to slightly perturbed values of $y_i$. Consider $\tilde{x}_i=(x_i + \Delta x_i)$, which is different from $x_i$ at $p$ locations.
\begin{align} 
U_{outer}(x_i + \Delta x_i) &= \sum_{j=1}^{N_{patt}} y_j x_j^T (x_i + \Delta x_i) \nonumber \\
&= (M-2p)y_i + \sum_{j\neq i} y_j x_j^T (x_i + \Delta x_i)
\end{align}
Note that $(x_i + \Delta x_i)$ is also a random vector and thus $x_j^T (x_i + \Delta x_i)$ can be considered a random variable with mean $0$ and variance $M$, which we will call $\tilde{X}^{prod}_{ij}$. The $k$-th entry of $U_{outer}\tilde{x}_i$ is
\[[U_{outer}\tilde{x}_i]^k = (M-2p)y^k_i + \sum_{j \neq i}y^k_j \tilde{X}^{prod}_{ij},\]
and following a similar argument to that above, the sum is a random variable with mean $0$ and standard deviation $\sqrt{\phi}M$, which we call $Z^k$. $\tilde{y}_i=U_{outer}\tilde{x}_i$ will differ from $y_i=U_{outer}x_i$ when flucutations in $Z^k$ exceed $(M-2p)$ (assuming $y_i^k = 1$ and signs reversed if not). This has a small constant probability at each node and thus $y_i$ and $\tilde{y}_i$ will differ on a small fraction of nodes.

\subsection*{Pseudoinverse construction for $U$}
We wish $U$ to solve the equation $U_{pinv}X \approx Y$, where $X$ and $Y$ are the matrices whose columns are the patterns $x_i$ and $y_i$. If $N_{patt}\leq M$, then this can be solved exactly\footnote{Unless the states are degenerate but this has vanishing probability} by choosing $U_{pinv}=Y(X^TX)^{-1}X^T$. Setting $q_i^T$ to be the i-th row of $(X^TX)^{-1}X^T$, we can write $U_{pinv}$ in a form analgous to the outer product construction as $U_{pinv} =\sum_{j=1}^{N_{patt}} y_j q_j^T$. 

Note that here $q_j^Tx_i = 0$ when $j\neq i$, so we avoid the additional error terms in the outer-product construction. However this requires each $q_i$ to be chosen considering all of the input patterns; thus we give up online learning. 

If we assume that the $q_j$'s are only weakly correlated with $(x_i + \Delta x_i)$ (for $i\neq j$), then a similar argument as above shows that slightly perturbed inputs are mapped to slightly perturbed outputs.

\bibliography{chaudhuri_master}